\documentclass[preprint2,12pt]{emulateapj}
\usepackage[]{graphicx}
\usepackage{amsmath}
\usepackage{amssymb}
\usepackage{mathrsfs}
\usepackage{subfigure}
\usepackage{boxedminipage}
\usepackage{aas_macros} 
\usepackage{natbib}
\usepackage{threeparttable}
\usepackage{appendix}
\usepackage{color,setspace,latexsym,amsthm,subfigure,rotating,longtable,arcs}
\usepackage{setspace}
\usepackage{Times}

\newcommand{\myr}{${\rm M_{\sun}\,yr^{-1}}$}
\newcommand{\Msol}{${\rm M_{\sun}}$}
\newcommand{\Lsol}{${\rm L_{\sun}}$}
\newcommand{\um}{$\mu$m}
\newcommand{\uJy}{$\mu$Jy}
\newcommand{\asec}{$^{\prime\prime}$}

\shorttitle{A Kinematic Study of Infrared Galaxies in a $z\sim0.9$ Cluster}
\shortauthors{Noble et al.}

\begin{document}

\title{A Kinematic Approach To Assessing Environmental Effects: Star-Forming Galaxies in a $\lowercase{z}\sim0.9$ S\lowercase{p}ARCS cluster using \textit{Spitzer} 24\,\um\ Observations}

\author{A.G. Noble\altaffilmark{1}, T.M.A. Webb\altaffilmark{1}, A. Muzzin\altaffilmark{2}, G. Wilson\altaffilmark{3}, H.K.C. Yee\altaffilmark{4}, R.F.J. van der Burg\altaffilmark{2}}

\altaffiltext{1}{Department of Physics, McGill University, 3600 rue University, Montr\'{e}al, Qu\'{e}bec, H3A 2T8, Canada}
\altaffiltext{2}{Leiden Observatory, Leiden University, PO Box 9513, 2300 RA Leiden, The Netherlands}
\altaffiltext{3}{Department of Physics and Astronomy, University of California, Riverside, CA 92521}
\altaffiltext{4}{Department of Astronomy and Astrophysics, University of Toronto, 50 St George Street, Toronto, Ontario M5S 3H4, Canada}

\begin{abstract}
We present an infrared study of a $z=0.872$ cluster, SpARCS\,J161314+564930, with the primary aim of distinguishing the dynamical histories of spectroscopically confirmed star-forming members to assess the role of cluster environment.  We utilize deep MIPS imaging and a mass-limited sample of 85 spectroscopic members to identify 16 24\,\um-bright sources within the cluster, and measure their 24\,\um\ star formation rates (SFRs) down to $\sim6$\,\myr.  Based on their line-of-sight velocities and stellar ages, MIPS cluster members appear to be an infalling population that was recently accreted from the field with minimal environmental dependency on their star formation.  However, we identify a double-sequenced distribution of star-forming galaxies amongst the members, with one branch exhibiting declining specific SFRs with mass.  The members along this sub-main sequence contain spectral features suggestive of passive galaxies.  Using caustic diagrams, we kinematically identify these galaxies as a virialized and/or backsplash population.  Moreover, we find a mix of dynamical histories at all projected radii, indicating that standard definitions of environment (i.e., radius and density) are contaminated with recently accreted interlopers, which could contribute to a lack of environmental trends for star-forming galaxies.  A cleaner narrative of their dynamical past begins to unfold when using a proxy for accretion histories through profiles of constant $(r/r_{200})\times(\Delta v/\sigma_v)$; galaxies accreted at earlier times possess lower values of  $(r/r_{200})\times(\Delta v/\sigma_v)$ with minimal contamination from the distinct infalling population.  Therefore, adopting a time-averaged definition for density (as traced by accretion histories) rather than an instantaneous density yields a depressed specific SFR within the dynamical cluster core.
\end{abstract}


\section{Introduction}
\label{sec:intro}
A hallmark of cosmology has been the establishment of hierarchical formation \citep{White91} in which growth of structure in the Universe proceeds via a ``bottom-up" scenario: matter condenses into low mass haloes which eventually merge to form larger structures, culminating with galaxy clusters.  Clusters subsequently evolve through the accretion of galaxies and groups along cosmic filaments.  This continual build-up of the cluster gives rise to distinct populations: a virialized component of older galaxies, and a younger population that was recently accreted from the surrounding low-density field \citep{Balogh98, Ellingson01}.  As clusters are thought to be hostile environments, member galaxies are exposed to various mechanisms that could potentially suppress their star formation, including ram-pressure stripping \citep[e.g.,][]{Gunn72}, strangulation \citep[e.g.,][]{Larson80}, and galaxy harassment, perhaps preceded by an initial burst \citep[e.g.,][]{Barnes91, Moore98}.  These mechanisms should induce marked differences in the young and in-situ population, as the latter has long endured the harsh conditions of the dense cluster.

Indeed, the local environment in which a galaxy resides is known to strongly correlate with several galaxy properties, such as star formation rate, stellar mass, color, and morphology.  Extensive observational efforts at low redshift have yielded a paradigm for galaxy dependencies, such that the densest regions at $z\sim0$ are devoid of star formation activity \citep[e.g.,][]{Gomez03, Balogh04a} and instead harbor massive \citep[e.g.,][]{Kauffmann04}, red \citep[e.g.,][]{Balogh04b, Hogg04, Baldry06},  early-type passive galaxies \citep{Dressler80}.  

Recently, the nature of these correlations at higher redshift, $z\sim1$, has become a rather contentious issue.  While it is well established that higher redshift clusters contain increased star formation activity compared to their local counterparts (e.g., \citealp{BO78, BO84, Ellingson01, Saintonge08}, Webb et al.\ in preparation) in parallel to the rapid decline of the cosmic star formation rate (SFR) since $z\sim1$ \citep[e.g.,][]{Lilly96, Madau96, Lefloch05}, there has not yet been a clear convergence on the SFR-density relation at this epoch.  Some studies have observed a reversal of the $z\sim0$ relation such that the predominant sites of star formation have migrated to denser regions by $z\sim1$ \citep{Elbaz07, Cooper08, Li11}, while many other groups find the local relation is already in place in  $z\sim1$ clusters (e.g., \citealp{Patel09, Patel11, Koyama10, Muzzin12}, Webb et al.\ in preparation).  These discrepancies most likely stem from differing selection effects (stellar mass versus luminosity limited samples) and the varying degrees of densities probed (cluster versus group environments).

Moreover, the situation is muddled by the interdependence between environment and stellar mass---whether these correlations with environment are causal or incidental is still ambiguous, since massive galaxies preferentially reside in dense regions.  Indeed, many properties seem to be also governed by stellar mass, including color \citep[e.g.,][]{Bell01, Kauffmann03} and SFR \citep[e.g.,][]{Brinchmann04, Noeske07}.  As surveys have attempted to untangle this covariance between stellar mass and environment, a host of trends have been revealed to depend unilaterally on one, or equally on each property.  For instance, both mass and environment have a separable effect on the fraction of star forming galaxies as seen by a decline in star-forming galaxies with increasing density and mass while the other parameter remains fixed \citep[e.g.,][]{Poggianti08, Peng10, Sobral11, Muzzin12}.  In contrast, many studies have found that the specific star formation rate (SSFR; the star formation rate per stellar mass) for star-forming galaxies is correlated with stellar mass in different environments but fails to exhibit any dependence on the local environment \citep[e.g.,][]{Kauffmann04, Poggianti08, Peng10, Lu12, Muzzin12}.  A possible explanation for this discordance that has been adopted by many groups is a rapid quenching timescale that suddenly alters the host galaxy's star-forming classification (e.g., color) preceding a decline in the observed SSFR \citep{Peng10, Muzzin12, Wetzel12_bimod}.	

Alternatively, a flat trend in SSFR with environment could suffer from radial projection effects.  If radial/density bins harbor a mixture of both physically high and low radius galaxies, any correlation with star forming properties could potentially get washed out in projection space.  Moreover, this could preferentially affect star forming galaxies over the quiescent population if they are inherently more biased towards radial contamination.  This is similar to the rationale put forward by \cite{Patel11}, who attributed a declining SFR-density (over all quiescent and star-forming galaxies) at $0.6<z<0.9$ to a combination of two effects: a varying fraction of passive and active galaxies with density, and suppressed SFRs at higher densities.

The recent work of \cite{Haines12} provides further insight into this contamination scenario via the accretion histories of cluster galaxies.  Utilizing caustic diagrams from the Millennium Simulations, \cite{Haines12} effectively isolate active galactic nuclei in projected velocity/radius space and determine that they are primarily an infalling population in spite of their low projected radii.  Applying the results of accretion histories from their work could potentially incite the emergence of an alternative environmental trend for star-forming galaxies that relies on more of a time-averaged rather than an instantaneous definition of density.

Here, we present an infrared study of a $z=0.872$ galaxy cluster drawn from the Spitzer Adaptation of the Red-sequence Cluster Survey \citep[SpARCS;][]{Wilson09, Muzzin09, Demarco10}.  We utilize extensive optical spectroscopy \citep{Muzzin12} and deep \textit{Spitzer}-MIPS 24\,\um\ observations to pinpoint dusty star-forming cluster galaxies.  By adapting a kinematic approach to the star formation histories of these galaxies via caustic diagrams, we intend to alleviate some of the confusion in environmental trends caused by radial projection effects.  

The paper is outlined as follows.  In \S\ref{sec:obs} we present the SpARCS survey and 24\,\um\ observations.  We briefly describe our computation of stellar masses and star formation rates in \S\ref{sec:analysis}.  Our results are presented in \S\ref{sec:results}, including stellar age, kinematic, and star formation trends.  We discuss the implications of these trends in \S\ref{sec:discussion}, and introduce a kinematic approach to classifying the star formation histories of cluster galaxies, along with a new interpretation of the environmental dependence on the SSFR of star-forming galaxies.  We conclude in \S\ref{sec:conclusions}.  Throughout the paper we assume a cosmology with $H_0=70$\,km\,s$^{-1}$\,Mpc$^{-1}$, $\Omega_{\textup{M}}=0.3$, $\Omega_{\Lambda}=0.7$.  Stellar masses and SFRs are based on a Chabrier initial mass function \citep{Chabrier03}.

\section{Observations and Data Reduction}
\label{sec:obs}

\subsection{The SpARCS Survey}
\label{sec:sparcs}
The Spitzer Adaptation of the Red-sequence Cluster Survey (SpARCS) is a $\sim45$ sq.\ deg.\ survey with deep $z^\prime$-band imaging from the CFHT and CTIO, designed to produce a large, homogeneously selected sample of massive galaxy clusters at $z > 1$.  By combining the $z^\prime$-passband observations with IRAC imaging from the Spitzer Wide-area InfraRed Extragalactic (SWIRE) survey,  SpARCS has discovered high-$z$ massive cluster candidates using either the red-sequence method \citep{Gladders00, Muzzin08} or the stellar bump sequence method \citep{Muzzin13}.  With $\sim200$ massive cluster candidates, including $\sim12$ $z > 1$ spectroscopically confirmed clusters \citep{Wilson09, Muzzin09, Demarco10}, SpARCS currently has one of the largest repositories of $z > 1$ galaxy clusters.  

\subsection{SpARCS J161314+564930  and the GCLASS Cluster Sample}
\label{sec:gclass}
This work presents an infrared study of SpARCS J161314+564930, a rich galaxy cluster at $z=0.872$ from the Gemini Cluster Astrophysics Spectroscopic Survey \citep[GCLASS;][]{Muzzin12} that was discovered using the red-sequence method.  It is an extremely massive cluster with: a velocity dispersion of $1350\pm100$\,km\,s$^{-1}$; $r_{200}$, the radius at which the cluster density is 200 times the critical density, of $2.1\pm0.2$\,Mpc; and M$_{200}=26.1^{+6.3}_{-5.4}\times10^{14}\,$\Msol (Wilson et al.\ in preparation).  This cluster field has $\sim$180 spectroscopic redshifts from GMOS-N on Gemini.  Galaxies within 3375\,km\,s$^{-1}$ of the cluster velocity dispersion (2.5$\sigma_v$) are considered to belong to the cluster, yielding 95 confirmed cluster members.  As GCLASS is a 3.6\,\um-selected survey, the spectroscopic data is rest-frame $H$-band selected and therefore close to a stellar mass-selected sample.  The final sample includes 85 cluster galaxies above our mass completeness limit of $2.0\times10^{9}\,$\Msol\ (see \S\ref{sec:mass}).  

\subsection{Spitzer-MIPS Imaging}
\label{sec:mips}
Our primary data set derives from the Multiband Imaging Photometer for \textit{Spitzer} (MIPS; \citealp{Rieke04}) aboard the \textit{Spitzer Space Telescope} \citep{Werner04} at 24\,\um\  with the aim to detect any dusty emission associated with the spectroscopically confirmed members.  With a 5.4 sq.\ arcmin field of view, MIPS provides sufficient coverage of the entire spectroscopic area.  The observations were part of the Guaranteed Time Observer program and completed in 2008 (proposal ID 50161) with an exposure of 1200 seconds per pixel. The MIPS image was reduced using a combination of the Spitzer Science Center's MOPEX software and a suite of IDL routines we developed to further optimize background subtraction (Webb et al.\ in preparation will provide more details).

\subsection{Source Detection and Photometry}
\label{sec:phot}
\begin{figure}[h!]     
\centering
   \includegraphics[width=9cm]{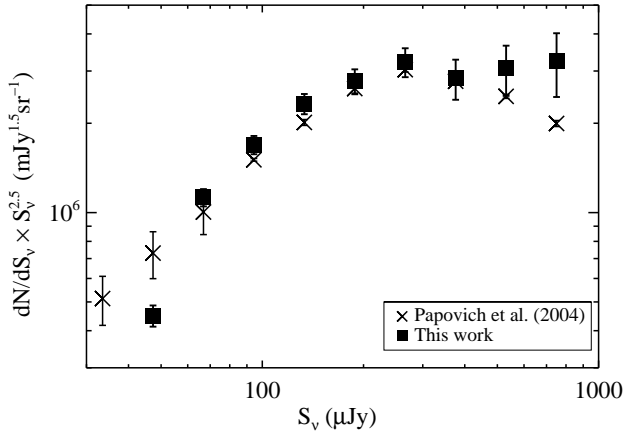} 
   \caption{The 24\,\um\ differential number counts of our catalog, normalized to the Euclidean slope (filled squares).  We compare to published counts from \cite{Papovich04} to determine a completeness limit of 70\,\uJy.  }
   \label{fig:counts}
\end{figure}

A source catalog of the entire MIPS field contains flux densities estimated with DAOPHOT \citep{Stetson87}, and positions using a photometry pipeline \citep{Yee91}.  We plot the 24\,\um\ differential number counts for our catalog in Figure~\ref{fig:counts}; they are in good agreement with the published 24\,\um\ counts from \cite{Papovich04}.  We determine the catalog completeness limits by locating the flux at which our counts deviate by $>2\sigma$ from the expected value of \cite{Papovich04}, which occurs at fluxes below $\sim$70\,\uJy.  This corresponds to an infrared luminosity of $\sim3\times10^{10}\,$\Lsol\ at $z=0.872$.

\section{Analysis}
\label{sec:analysis}

\subsection{Counterpart Identification}
\label{sec:false}
We search for optical counterparts within $\sim$2\asec\ to locate cluster members with 24\,\um\ emission; we detect 16 MIPS cluster members which constitute the focus of our study.  To quantify the number of chance alignments between 24\,\um\ and optical sources, we drop 2\asec\ apertures at random positions on the MIPS field and count the number of times we find one or more source.  We perform 100 realizations at each flux level within 40--400\,\uJy\ in 10\,\uJy\ steps.  At 70\,\uJy, the completeness depth of our MIPS catalog, we expect 5\% of MIPS counterparts to be falsely identified.  This translates into one chance alignment in our counterpart sample of 16 MIPS cluster members.

In order to assess the effect of the cluster environment on its constituent galaxies, it is essential to have a control sample of field galaxies that are distinct from the cluster.  Our spectroscopic campaign for the entire GCLASS survey successfully obtained redshifts for a significant number of foreground and background sources in the fields of the clusters with identical selection effects as cluster galaxies; these sources constitute our field sample (see \citealt{Muzzin12}).  We employ the same criteria to search for MIPS counterpart emission to the field and cluster galaxies, but restrict the field redshift range to $0.84<z<1.0$ in order to probe the same epoch as the $z=0.872$ cluster. This leaves us with ten 24\,\um-bright field galaxies above our flux limit.

\subsection{Stellar Masses}
\label{sec:mass}
We compute stellar masses using a technique similar to \citet{Bell01} and \citet{Kauffmann03}.  In particular, we utilize a spectral feature that provides information on the age of the stellar population but is fairly insensitive to dust, the 4000-\AA\ break. Using the \citet{Bruzual03} stellar population synthesis models with solar metallicity, assuming a Chabrier IMF, and adopting a star formation timescale of $\tau=0.3$\,Gyr, we infer a M/L ratio at 3.6\,\um\ from the strength of the 4000-\AA\ break.  We than convert to a stellar mass using the measured 3.6\,\um\ luminosity (for more details see \citealp{Muzzin12}).   In order to obtain a mass-limited sample, we do not include sources below our mass completeness limit of $2\times10^{9}\,$\Msol\ \citep{Muzzin12}.

\subsection{24\,\um\ Star Formation Rates}
\label{sec:sfrs}

The mid-infrared luminosity of a galaxy probes thermal emission from dust grains that has been reprocessed from UV light, and therefore offers a clean measurement of the star formation rate (SFR) that is minimally affected by dust extinction \citep{Kennicutt98}.  Moreover, rest-frame 24\,\um\ traces dust heated by younger stars and therefore provides a measure of the instantaneous SFR \citep{Calzetti07}.  However, at $z=0.872$, 24\,\um\ corresponds to 13\,\um\ rest-frame, which has recently been shown to contain a higher contribution of dust heated from intermediate age stars,  which is evidence for star formation over longer timescales ($\sim$1--2\,Gyrs; \citealp{Salim09}).  As there is a rapid decline in the cosmic star formation since $z\sim1$ \citep{Lilly96, Madau96, Lefloch05}, this could result in an overestimate of the star formation as derived from observed 24\,\um\ flux.  In parallel, \cite{Rodighiero10} found that extrapolating an infrared luminosity from the monochromatic 24\,\um\ flux leads to an underestimate by a factor of $\sim1.6$ compared to Herschel measurements observed at 100 and 160\,\um\ for $0.5<z<1.0$ galaxies.  Despite some potential systematic effects in the 24\,\um-derived SFRs, the results from this study are minimally affected, as we rely solely on relative differences within our own sample (see also \citealp{Patel11}).

We calculate the 24\,\um-derived SFR by converting the MIPS flux into a total infrared luminosity using an average of the \citet{Chary01} and \citet{Dale02} templates.  We employ the relation in \citet{Kennicutt98} to calculate a SFR and convert to a Chabrier-IMF based SFR using a factor of 1.65.   Based on the depth of our MIPS images, we probe down to $\sim6\,$\myr.

\subsection{Active Galactic Nucleus Contamination}
\label{sec:sfrs}

We assume that the MIPS sources are dominated by star formation, with little or no contamination from active galactic nuclei (AGN); this seems reasonable given that only $\sim5$\% of infrared field galaxies at $z\sim1$ have their total IR luminosity dominated by AGN emission \citep{Fadda10}, and the contribution declines drastically with low mid-infrared flux.  In fact, AGNs only begin to dominate the infrared emission for 24\,\um\ fluxes above 1.2\,mJy at $z\sim0.8$, and are missing below 0.8\,mJy \citep{Fu10}.  We note that our sample contains only two sources at the high flux end, at 0.83\,mJy and 0.91\,mJy, both of which are still below the level where AGNs prevail.  Moreover, \cite{Martini09} measured the occurrence of AGNs in cluster galaxies and found that while the AGN fraction increases with redshift, it is still only 1.5\% at $z=0.81$.  

Regardless, we have performed two quick checks for obvious AGN sources.  We first compare the equivalent width of the [OII] line and the depth of the 4000-\AA\ break, following \cite{Stasinska06}.  Our sample contains ten MIPS sources with measurable [OII] emission, all of which are all consistent with the normal star forming galaxies shown in fig.~7 from \cite{Stasinska06}.  We also exploit infrared color diagnostics \citep[e.g.,][]{Lacy04, Sajina05} using data from the Infrared Array Camera (IRAC; \citealp{Fazio04}).   All but two MIPS cluster members display infrared colors consistent with stellar- and PAH-dominated sources (i.e., star-forming) at $z\sim0.9$.  The other two sources lie on the edge of the region encompassed by AGN, but we note there is a high contamination from PAH-dominated sources at this location; these are the same two sources described previously with high mid-infrared fluxes.  In the absence of any concrete evidence for pure AGNs, we assume the MIPS flux is dominated by star formation.

\section{Results}
\label{sec:results}
This analysis presents an infrared study of a single GCLASS cluster at $z=0.872$.  As such, we focus primarily on the properties of the MIPS population and utilize the extensive spectroscopy for verification of MIPS cluster membership and an estimate of stellar age.  We refer the reader to \citet{Muzzin12} for a detailed description of the spectroscopic selection criteria and completeness, as well as a comprehensive analysis of the GCLASS clusters as seen through optical spectroscopic measures, for example, [OII] star-formation rates.

\subsection{Age as a Function of Stellar Mass}
\label{sec:d4000}

\begin{figure}[h!]    
 \centering 
   \includegraphics[width=9cm]{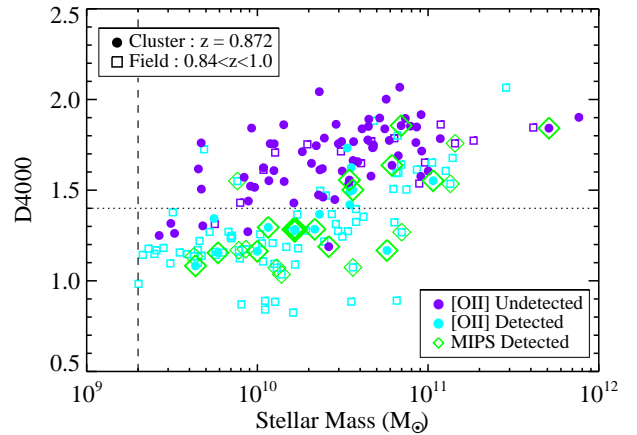} 
      \caption{The strength of the 4000-\AA\ break as a function of stellar mass for the cluster ($z=0.872$; filled circles) and coeval field ($0.84 < z < 1.0$; open squares) population.  The galaxies with [OII] emission (star forming) are plotted in cyan, while those without any [OII] emission are purple.  A green diamond denotes coincident 24\,\um\ emission; there are three MIPS detections, all with [OII] emission, at the location of the thickest green diamond with a D4000 of 1.3 and stellar mass of $\sim1.7\times10^{9}\,$\Msol.  The dashed vertical line illustrates our mass limit of $2.0\times10^{9}\,$\Msol, and the dotted horizontal line indicates our cut in D4000 in subsequent plots.  We note that the two most massive cluster members have both been identified as the brightest cluster galaxies, although only one is detected at 24\,\um.}
   \label{fig:d4000}
\end{figure}

Here we investigate the depth of the 4000-\AA\ break (D4000) as a function of stellar mass for the MIPS-detected cluster members compared to the larger sample of spectroscopic members.  The 4000-\AA\ break, defined as the ratio of the integrated flux density at 4000\,\AA--4100\,\AA\ to that blue-ward of the break, 3850\,\AA--3950\,\AA\ \citep{Balogh99}, arises from an accumulation of absorption lines and increases in depth for old and metal-rich stellar populations.  It can therefore be used as a proxy for stellar age.  In Figure~\ref{fig:d4000} we plot the depth of the 4000-\AA\ break for all cluster galaxies (filled circles), as well as a field population over the same epoch, $0.84<z<1.0$ (open squares).  We highlight cluster and field galaxies detected at 24\,\um\ with green diamonds.  The MIPS cluster galaxies are primarily coincident with the coeval field population, spanning the same range of stellar mass and D4000.  A Kolmogorov-Smirnov test between the MIPS cluster members and general field population reveals a 76\% chance they derive from the same parent population, in stark contrast to that of cluster members without MIPS detections, where the null hypothesis is rejected at a high significance of 0.02\%.  Compared to the spectroscopically-confirmed members not detected at 24\,\um, MIPS members form a young envelope of the cluster population at a given mass, which suggests they were recently accreted from the field.

We also distinguish between galaxies with [OII] emission and those without (i.e., below $\sim1$ (3)\,\AA\ equivalent width for the highest (lowest) signal-to-noise spectra); in Figure~\ref{fig:d4000} cyan circles have measurable [OII] which is indicative of ongoing star formation, while purple circles represent quiescent (or extremely dust-enshrouded) galaxies.  The majority of MIPS galaxies are [OII]-detected in the cluster and field, albeit at a slightly reduced fraction in the cluster (70 and 90\%, respectively).  This is not surprising given the recent conclusion from Webb et al. (in preparation) that optical studies are not significantly biased against dusty star formation.  We define a D4000 limit of 1.4 (dotted line), above which $>90\%$ of cluster members without [OII] emission reside, signifying a minimum age for the older, quiescent cluster galaxies.  We note that a variable D4000 as a function of stellar mass does not significantly affect our conclusions, and therefore adopt the simpler flat delineation.  In all subsequent plots, we differentiate galaxies by their nominal age as traced by the strength of the 4000-\AA\ break, rather than [OII] detections as in Figure~\ref{fig:d4000}.

\subsection{Relative Velocity as a Function of Radius}
\label{sec:vel}
\begin{figure}[h!]     \centering
   \includegraphics[width=9cm]{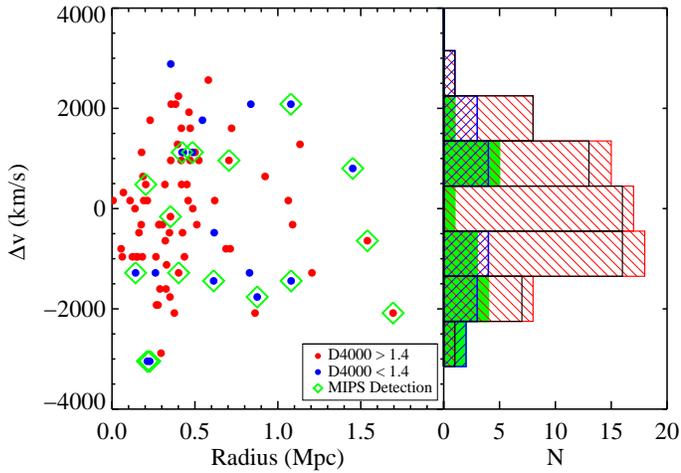} 
   \caption{The relative velocity as a function of clustercentric radius for all spectroscopic members.  Red circles corresponds to older galaxies with D4000 $>1.4$ and blue circles denote younger galaxies, D4000 $<1.4$.  The histograms correspond to the number of galaxies within 900\,${\rm km\,s^{-1}}$ velocity bins; the colors correspond to the same types of galaxies in the legend on the left.  The black histogram represents the population of old galaxies with the MIPS sources removed (6 in total). The green histogram shows all MIPS sources, regardless of age.}
   \label{fig:v_hist}
\end{figure}

In the left panel of Figure~\ref{fig:v_hist} we plot the relative velocity of all cluster members as a function of clustercentric radius, as defined by the projected distance to the brightest cluster galaxy (BCG), the brightest 3.6\,\um\ cluster member, which is also the most massive (see \citealp{Lidman12} for a detailed discussion of the BCG selection in SpARCS, including this cluster).  As discussed in \S\ref{sec:d4000}, we now separate the galaxies into old (red circles) and young (blue circles) cluster populations based on D4000, and highlight 24\,\um-bright members with green diamonds.  Barring two MIPS galaxies at low projected velocities and radii (both of which are classified as older), MIPS sources tend to display larger velocities on average ($\lvert\overline{\Delta v}\rvert = 1422\pm205.8$\,km\,s$^{-1}$), in contrast to a tighter distribution expected from a virialized population ($\lvert\overline{\Delta v}\rvert = 995.1\pm89.60$\,km\,s$^{-1}$ for older galaxies without a MIPS detection).  Therefore, the velocity-radial space encompassed by MIPS galaxies further supports the idea that they have been recently accreted.  

The histogram in the right panel of Figure~\ref{fig:v_hist} emphasizes this point: the older cluster population (red hashed histogram) displays a Gaussian distribution of velocities and primarily falls within the cluster velocity dispersion (1350\,km\,s$^{-1}$), while the young (blue hashed histogram) and MIPS (solid green histogram) galaxy populations exhibit bimodal and/or flat distributions, peaking at velocities greater than $\pm$1000\,km\,s$^{-1}$.  Moreover, the younger members within the cluster avoid the central velocity bin completely.  

Given that 38\% of MIPS cluster galaxies have a steep 4000-\AA\ break with D4000 $>1.4$ (red circles with green diamonds), we might expect to see some overlap in their velocities compared to those in the larger spectroscopic sample of older galaxies (i.e., the confirmed members that are not detected by MIPS; red circles without a green diamond).  However, a Kolmogorov-Smirnov test between the velocities of all MIPS galaxies (green diamonds) and older cluster galaxies (the black histogram shows old cluster members with old MIPS sources removed) rejects the null hypothesis with marginal significance---there is only a 6\% probability that the two distributions derive from the same parent population.  It therefore seems likely that the MIPS galaxies trace a younger, recently accreted population that is not yet virialized with the cluster.

\subsection{Environmental Dependence on the Specific Star Formation Rate}
\label{sec:ssfr}
Given the implication that MIPS galaxies represent a younger, infalling population of cluster members, we might expect to detect a correlation between the specific star formation rate (SSFR; star formation rate per stellar mass) of MIPS galaxies and local environment.  If the majority of MIPS galaxies do in fact belong to an infall population, there should be more star formation in the cluster outskirts compared to the cluster core, where galaxies have had their star formation shut off as they fall deep into the cluster.  

\begin{figure*}[]     \centering
   \subfigure{\includegraphics[width=9.8cm]{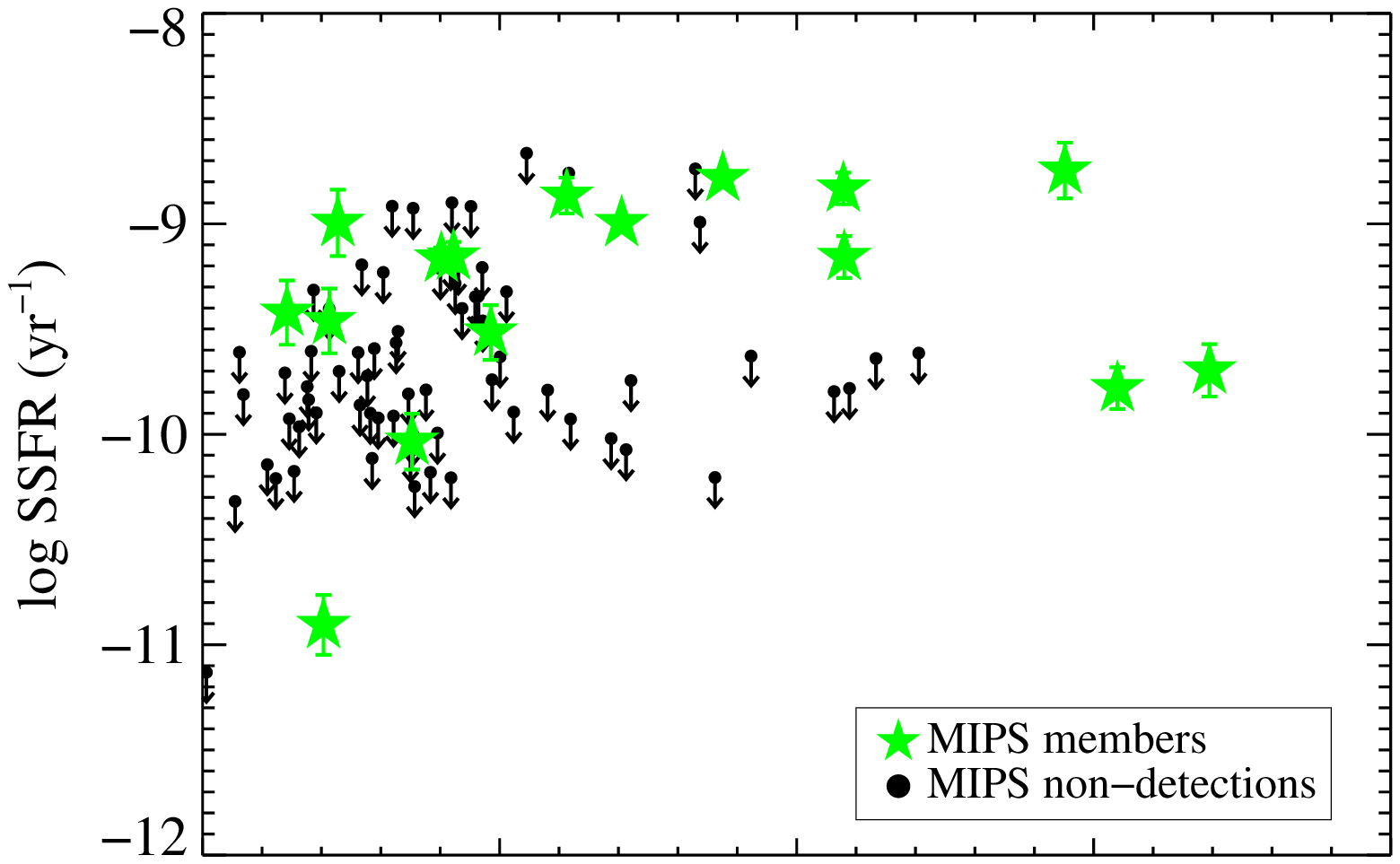} }\hspace{-2cm} \vspace{-1.5cm}
\subfigure{\includegraphics[width=9.8cm]{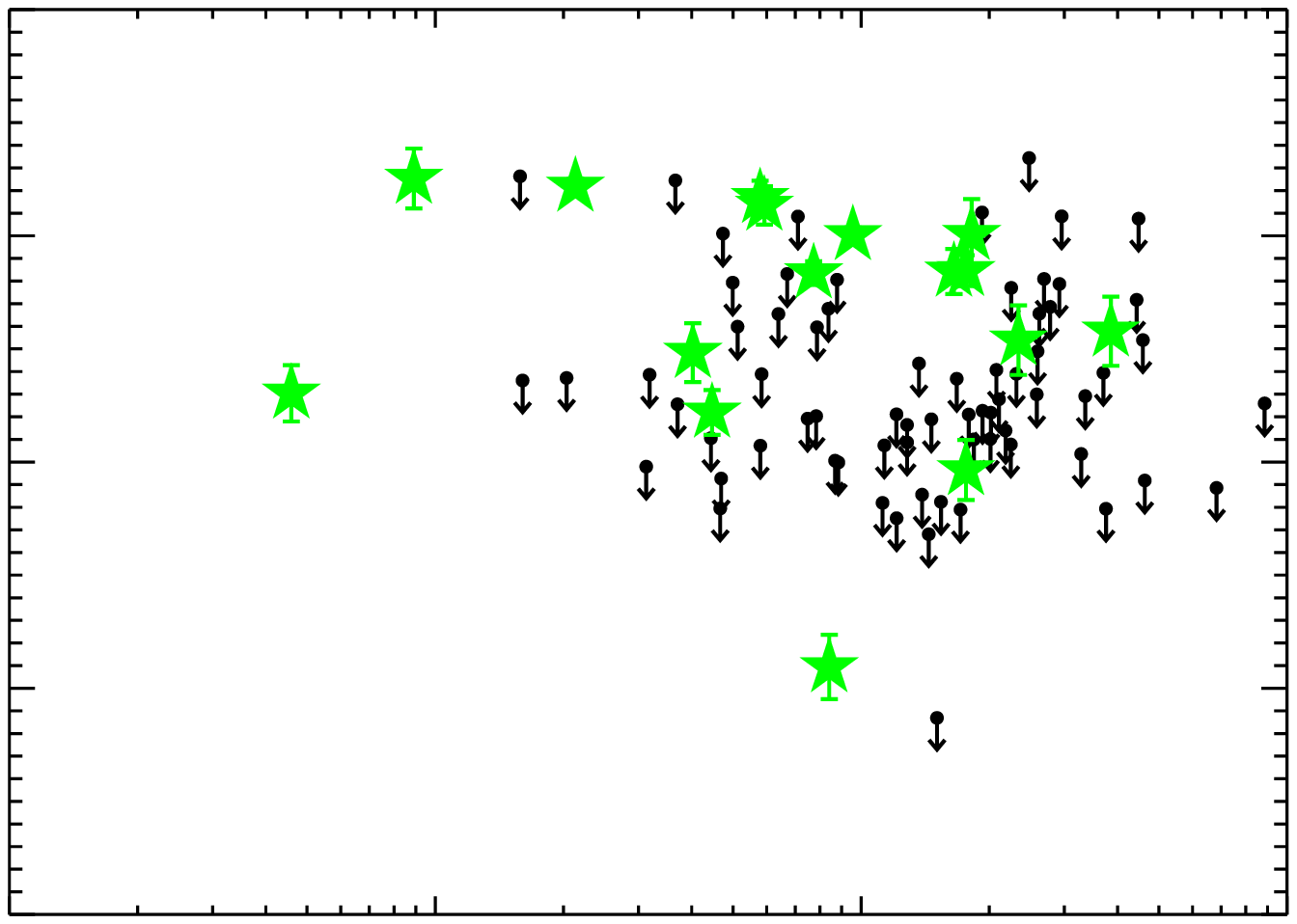} } 
\subfigure{\includegraphics[width=9.8cm]{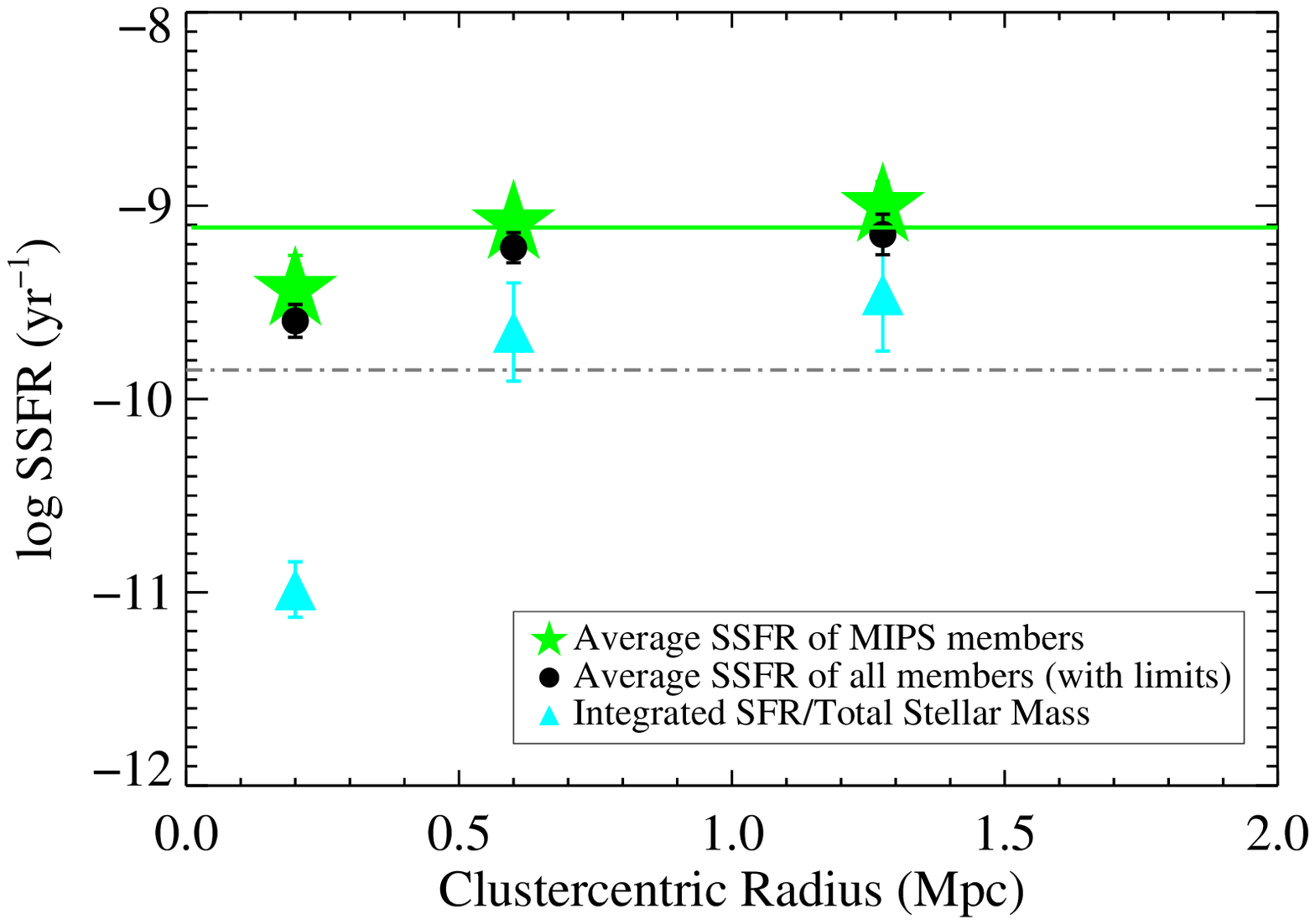} }\hspace{-2cm} 
\subfigure{\includegraphics[width=9.8cm]{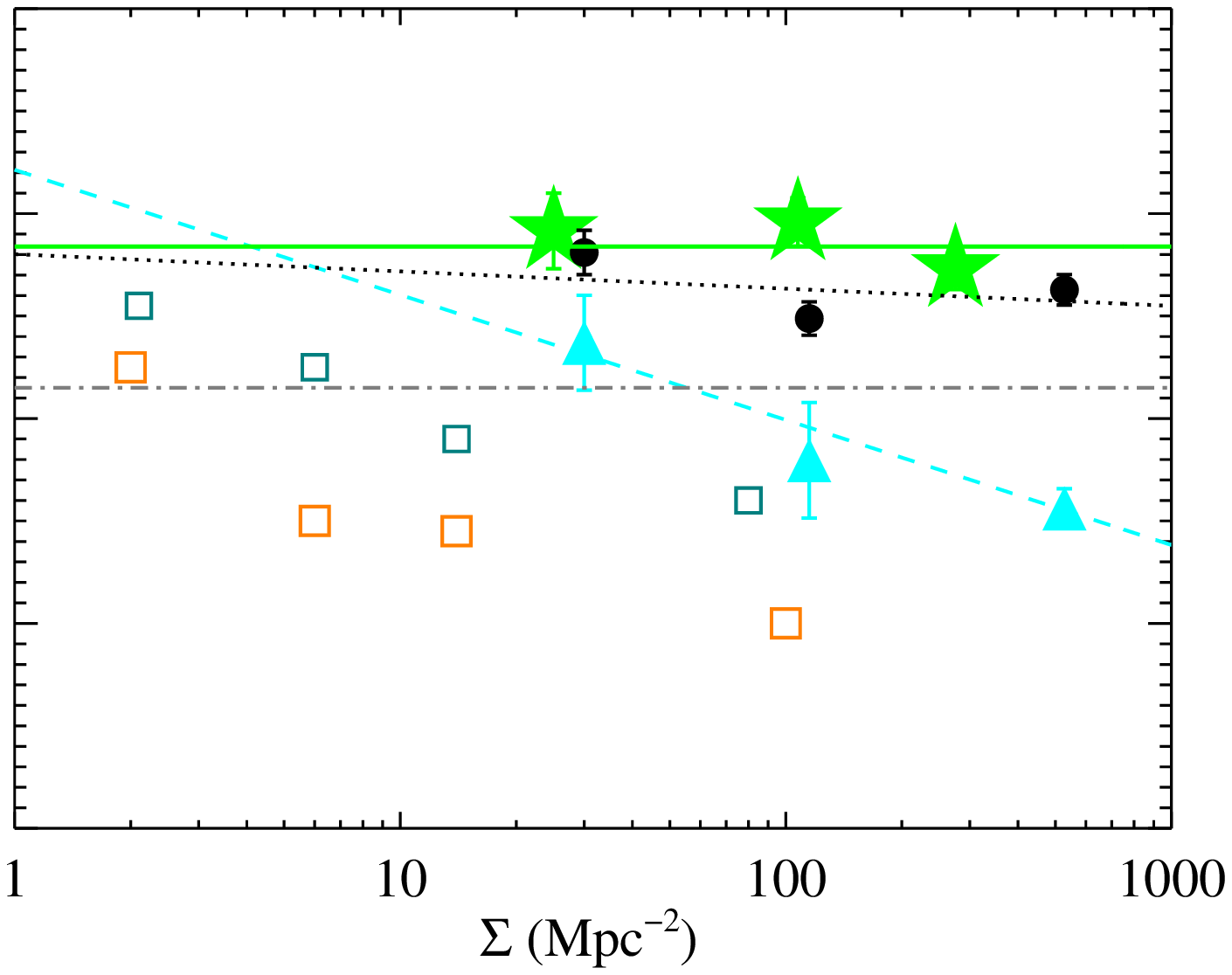} } 
   \caption{Upper panels---The individual SSFRs for all cluster members as a function of clustercentric radius (left panel) and density using the 10th nearest neighbor (right panel).  The green stars correspond to MIPS cluster members, and the black circles with arrows are 3$\sigma$ upper limits on the SSFR given the stellar mass for the cluster members without MIPS detections.  Lower panels---The SSFR in bins of clustercentric radius (left) and density (right).  The green stars correspond to the average SSFR of detected MIPS members in each bin, i.e., the SFR divided solely by the mass of MIPS members and averaged over the number of MIPS galaxies in each bin.  Black circles represent the average SSFR for all cluster members, assuming a 3$\sigma$ upper limit on the SSFR for undetected MIPS members.  Cyan triangles denote the integrated SFR (above our SFR-limit) per total unit stellar mass: a sum of the total 24um-derived SFR divided by the total stellar mass of all spectroscopic members in each radial bin, which essentially probes the fraction of star-forming galaxies.  The horizontal solid green line illustrates the best fit line to the green stars with the slope fixed to zero, i.e., the normalization of the average MIPS SSFR. The dot-dashed horizontal line corresponds to the required SSFR for the mass to double by $z=0$.  The two remaining lines in the lower right panel depict the best linear fits to the corresponding binned values with density.  The open squares represent the stacked IR SSFRs from \citet{Patel09} in a $z=0.834$ cluster, where teal squares correspond to galaxies with masses of $2.0\times10^{10} < $ M/\Msol $< 6.3\times10^{10}$, and the grey squares represent all galaxies with M $>6.3\times10^{10}$\,\Msol.}
   \label{fig:ssfr_rad}
\end{figure*}

In Figure \ref{fig:ssfr_rad}, we plot the SSFR as a function of local environment, with binned averages shown in the lower panels.  The upper panels show the SSFR for every MIPS member (green stars), as well as an upper limit for every spectroscopic member, determined from the 3$\sigma$ completeness limit of the SSFR for a given stellar mass.  Errors on the averages are determined from 100 bootstrap resamplings of the data in each bin.  

We use two different proxies for environment in order to compare to previous studies, and to confirm we are not substantially biased with either parameterization.  In the left panels we show clustercentric radius, where the core of the cluster is defined by the location of the BCG.  The right panels contain local galaxy density calculated from the distance to the 10th nearest neighbor.  Although in the literature these two parameters are often used interchangeably, they sample different mechanisms: the former is a better indicator of the global environment and cluster potential, whereas the local density has a proximate effect on galaxies.  However, both parameters still play a significant role in the properties of cluster galaxies, and we therefore investigate trends with each separately (see \citealp{Li12} for a detailed discussion on the two methods).   While both measurements suffer from projection effects, the benefit of using local density is that it is straight-forward to correct each galaxy for spectroscopic completeness given its stellar mass and radius \citep{Muzzin12}, which essentially provides us with a 100\% complete sampling of density.  We note that our radial measurement does not require a completeness correction as the completeness bias for mass is similar at all radii (see fig.~4 in \citealp{Muzzin12}).  To compute density, each galaxy is first given a weight based on its completeness.  We then interpolate the distance to the 10th nearest neighbor for each source by summing the weighted values for galaxies above our mass limit of $2.0\times10^9$\,\Msol, and compute the density as $\Sigma=10/(\pi d^2_{10}$).  The drawback of this method is that we lose information on smaller density scales.   

Nevertheless, our two environment parameterizations convey the same trend: the average SSFR of star-forming galaxies (i.e., MIPS galaxies; green stars) is mostly independent of environment (see lower panels of Figure~\ref{fig:ssfr_rad}).  A best-fit line to the average MIPS SSFR in radial and density bins reveals a slope consistent with zero at the level of $1.6\sigma$ ($-0.29\pm0.18$) and $1.2\sigma$ ($-0.23\pm0.20$), respectively.  In Figure~\ref{fig:ssfr_rad}, we show the best fit to the normalization of the SSFR with the slope fixed at zero.  Though seemingly surprising, this flat trend is consistent with other cluster surveys over many redshifts and using various star formation tracers, including UV studies at $0.16<z<0.36$ \citep{Lu12} and optical star formation indicators at $z\sim1$ \citep{Peng10,Muzzin12}.  

However, our MIPS sample is SFR-limited, and thus not complete in SSFR for any given mass except at the highest SSFR values, log(SSFR) $\gtrsim-9$.  The flat trend in SSFR might be the result of skimming the high SSFR galaxies off the top of a deeper correlation.  For example, if we assume that more massive galaxies reside at lower radii, probing deeper in SFR could preferentially bring down the average SSFR at low radii compared to outer radii, uncovering a non-flat SSFR trend.   We check for this effect by assigning a 3$\sigma$ upper limit for the SSFR of cluster members without a MIPS detection (black arrows in upper panels of Figure~\ref{fig:ssfr_rad}) and include these limits in the average SSFR (black circles in lower panels).  The upper panels of Figure~\ref{fig:ssfr_rad} reveal that there is no underlying mass segregation in the radial/density properties in the population not detected by MIPS (i.e., the limits are flat with environment), which suggests that a lower SSFR-limit would not necessarily result in a non-flat trend for star-forming galaxies galaxies.  Moreover, the slope of the average SSFR assuming all cluster members have some star formation flattens out even further with density ($-0.08\pm0.10$, dotted black line in lower right panel).

If we instead treat the galaxies without MIPS detections as quiescent (i.e., non-star forming) and investigate the integrated SFR (above our SFR-limit) per total unit stellar mass as a function of environment (cyan triangles), namely the total MIPS SFR divided by the total stellar mass of all cluster members in each bin, we find a decreasing SSFR with increasing (decreasing) density (radius).  We note that both spectroscopic and MIPS members have the same target selection, so they have the same completeness rates as a function of mass and radius.
The integrated SFR per total stellar mass is a proxy for the number of star forming galaxies compared to quiescent galaxies at each radius and reveals a lower fraction at denser environments.  A fit to the integrated SSFR versus density yields a slope of $-0.61\pm0.19$ (dashed cyan line), and is therefore inconsistent with zero at the $>3\sigma$ level. 

This depression of star formation in the cluster core is in accordance with local clusters \citep[e.g.,][]{Kauffmann04} and suggests that the SSFR-density relation is already established in the highest-density regions at $z\sim0.9$.  This is in contrast to $z\sim1$ field studies that found a reversal in the SFR-density relation \citep{Elbaz07, Cooper08}, although they were limited to lower density environments.  Instead, we compare our results to a similar IR study of a $z=0.834$ cluster from \cite{Patel09} who probe a wide-range of environments and utilize a 24\,\um\ stacking analysis on all cluster members to determine SSFRs.  They uncover a trend of decreasing SSFR with increasing density (open squares in the right panel of Figure~\ref{fig:ssfr_rad}), analogous to our total SSFR.  This relation prevails even when controlling for stellar mass: the open, teal squares correspond to lower masses of $2.0\times10^{10} < $ M/\Msol $< 6.3\times10^{10}$, and the open orange squares represent all galaxies with M $>6.3\times10^{10}$\,\Msol\ in the \cite{Patel09} sample.  We note the normalization of our trend is higher, but we probe down to an order of magnitude lower in stellar mass ($2.0\times10^{9}$\,\Msol); as lower-mass systems typically display slightly higher SSFRs (see \S\ref{sec:sfr} below), a higher SSFR in our data is expected.  

In Figure~\ref{fig:ssfr_rad} we also plot the SSFR required for stellar mass to double (assuming a constant SFR) by $z=0$ (dot-dashed horizontal line).  We note that the integrated SSFR per total stellar mass (cyan triangles) is consistent with or below this limit in all environments, signifying that a majority of cluster members (including quiescent galaxies) have already experienced most of their mass growth.  However, the average MIPS SSFR (and average SSFR for all members with limits) lie above this line: the star-forming galaxies have yet to undergo the bulk of their activity, assuming they can sustain their SFR.  Perhaps we are witnessing the primal growth stage of these galaxies, before they have properly assimilated into the cluster and been exposed to any environmental quenching.  Alternatively, if any of these galaxies have already experienced a suppression of their star formation, the quenching timescale must be rapid enough that any environmental trend with SSFR gets washed out: SSFRs are immediately terminated, precluding an intermediate phase.  In other words, why is there a lack of variation in the star-forming population despite a changing fraction of star-forming galaxies with environment?  We will return to this flat environmental trend in \S\ref{sec:discussion} and provide an alternative explanation.

\subsection{The Correlation between Star Formation Rate and Stellar Mass}
\label{sec:sfr}
\begin{figure}[h!]     \centering
 \subfigure{\includegraphics[width=9cm]{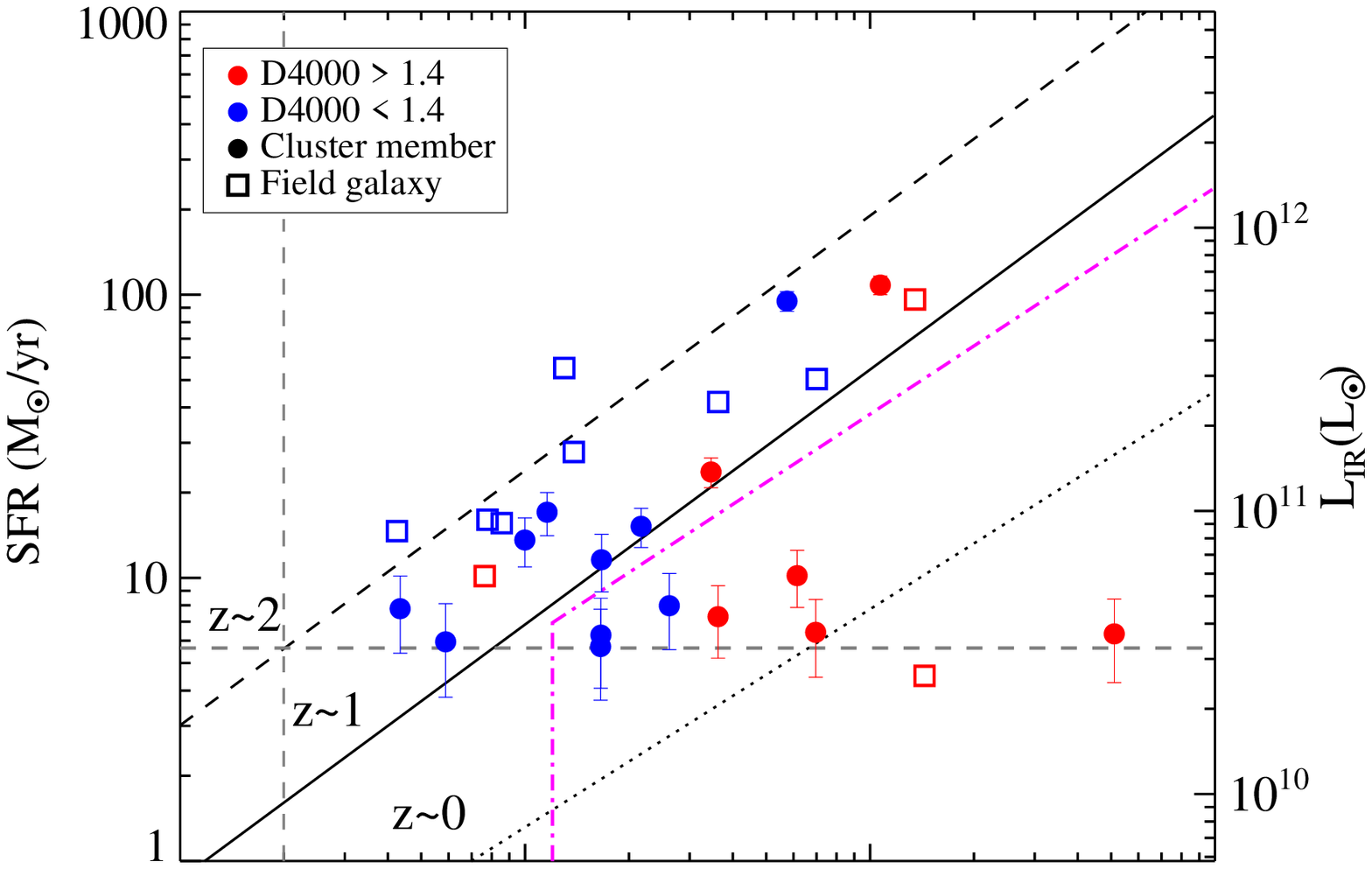}}
  \subfigure{\includegraphics[width=9cm]{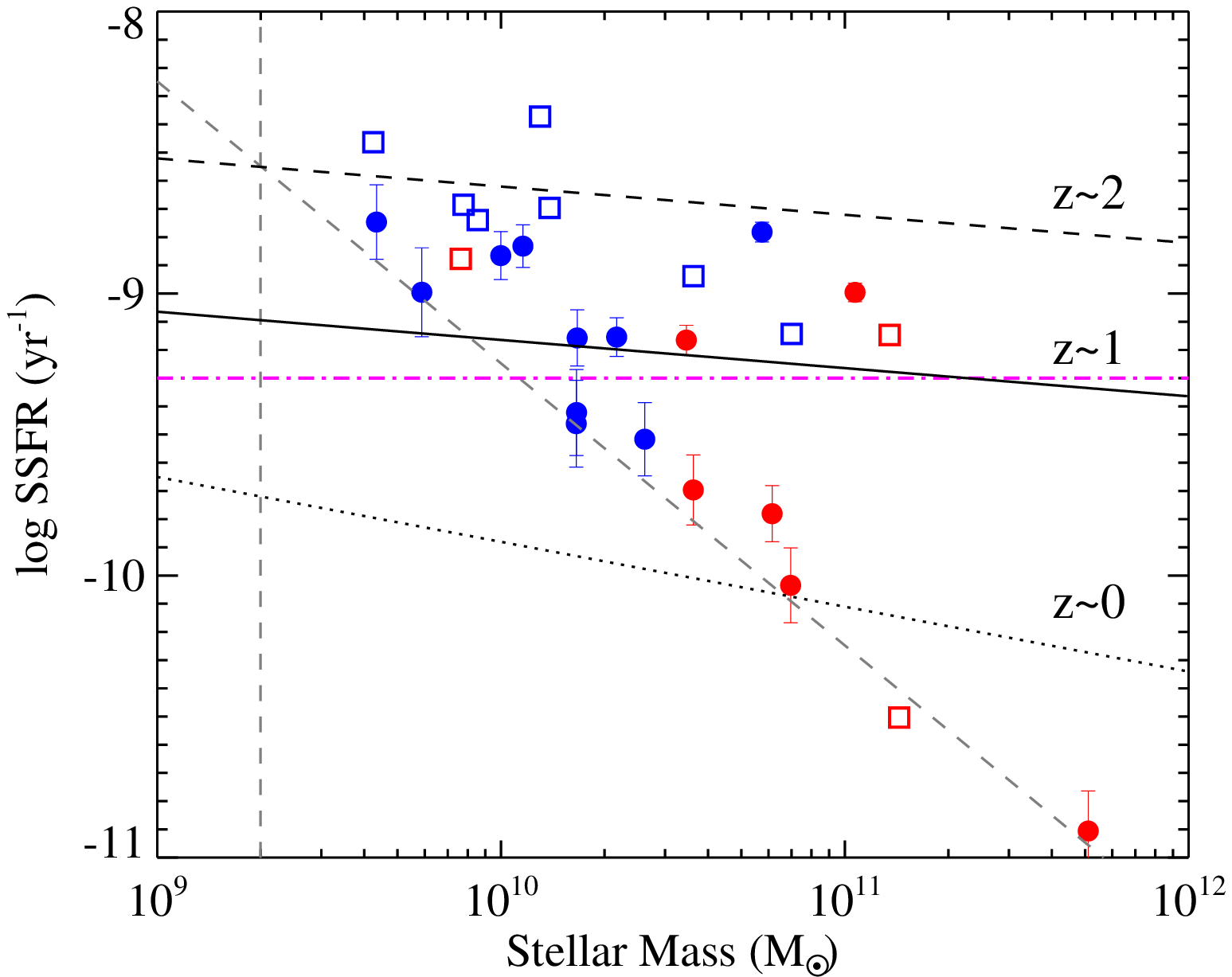}}
   \caption{Top---SFR versus stellar mass for MIPS cluster galaxies, separated in color by D4000.  Open squares are the field population from $0.84 < z < 1.0$.  The dotted, solid, and dashed black lines correspond to the field trends at $z=0,1,2$, respectively (converted to a Chabrier IMF).  The pink dot-dashed line denotes the boundary between the two populations of the cluster galaxies: main-sequence vs sub-main sequence, as defined by their star formation rate.  The vertical and horizontal gray dashed lines indicate our mass completeness, M $>2.0\times10^9$\,M$_{\odot}$, and SFR detection limit, respectively.  Bottom---SSFR versus stellar mass.  Symbols and lines are the same as the top panel.}
   \label{fig:sfr_mstell}
\end{figure}

We can investigate whether any MIPS members deviate from their expected SFRs and SSFRs given the tight correlation of increasing SFR with stellar mass for star-forming galaxies, which retains only 0.2 dex scatter at $z\sim1$.
Recent studies have observed this star-forming main sequence in the field out to $z\sim2$, which monotonically shifts to higher SFRs with increasing redshift  \citep{Noeske07, Elbaz07, Daddi07}.  In Figure~\ref{fig:sfr_mstell} we plot the main sequence trend in the field at various redshifts: $z=0.1$ from the Sloan Digital Sky Survey \citep[SDSS;][]{Brinchmann04} as analyzed by \citet{Elbaz07}; $z\sim1$ from the Great Observatories Origins Deep Survey \citep[GOODS;][]{Elbaz07}; and $z\sim2$ from GOODS \citep{Daddi07}; we have converted each trend to Chabrier-IMF based masses and SFRs for consistency with our own sample.  We also plot our 24\,\um-derived SFRs for MIPS cluster members at $z\sim0.9$ along with our field sample over $0.84<z<1.0$.  Due to systematic differences in SFRs, stellar masses, and selection criteria (e.g. mass-limited versus luminosity-limited samples), we refrain from quantitative comparisons between our trend and the field samples from GOODS and SDSS.  Our own field sample offers a more suitable comparison, and in fact, displays a similar trend as the larger GOODS sample at $z\sim1$, albeit with slightly higher SFRs.

Immediately obvious in Figure~\ref{fig:sfr_mstell} is the distinction between the old and young MIPS populations: the majority of massive, older members (filled red circles) lie well below the expected correlation, while the younger members (filled blue circles) tend to follow the main sequence trend observed in the $z\sim1$ GOODS field (solid black line) and the GCLASS field (open squares).  Moreover, this double-sequence in SFRs is unique to the cluster as there is only one field galaxy in our sample with a significantly low SFR for its stellar mass.

The double-branched distribution is similarly manifested in the SSFR, plotted in the lower panel of Figure~\ref{fig:sfr_mstell}.  The least massive galaxies, which are inherently younger (see Figure~\ref{fig:d4000}), have the highest SSFRs, while more massive galaxies have substantially lower SSFRs.  While this is consistent with the evolution of the mass function \citep[e.g.,][]{Kodama04} which shows that the massive end of the galaxy mass function in clusters is in place by $z\sim1$ and the evolution between $z = 0-1$ consists of a build-up of the $< 10^{11}$\Msol\ end, it could also be a completeness effect as we do not probe the region of low mass and low SSFR (Figure~\ref{fig:sfr_mstell}).

In hopes of identifying further differences, we split the 24\,\um\ members into two cases based on their proximity to the expected SFR for their given mass at $z\sim1$, as shown by the dot-dashed pink line in Figure~\ref{fig:sfr_mstell}.  This line corresponds to log(SSFR)$=-9.3$.  We designate the population above the line as the main sequence population, and those below the line as the sub-main sequence given they are in fact closer to the $z\sim0$ field trend and lie in a substantially disparate space from the GCLASS $z\sim1$ field sample. 

\begin{figure}[h!]    
 \centering
\includegraphics[width=9cm]{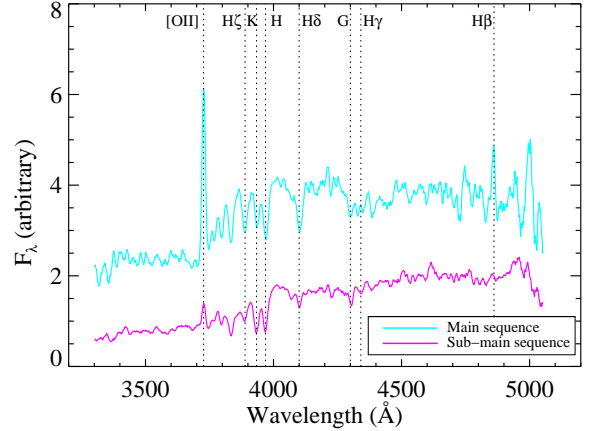}
   \caption{The resulting stacked spectra of MIPS cluster members from each population identified in \S\ref{sec:sfr}: main sequence population (cyan spectrum) and sub-main sequence population (pink spectrum). Individual spectra are weighted by their spectroscopic completeness, as determined by their stellar mass and clustercentric radius.} 
   \label{fig:pop_stacked}
\end{figure}

In Figure~\ref{fig:pop_stacked} we stack the spectra in each population separately (nine and seven members in the main sequence and sub-main sequence groups, respectively) with an inverse weighting based on the spectroscopic completeness.  Specifically, we investigate the equivalent widths of the [OII] doublet (3727\,\AA) and a Balmer absorption line, H$\delta$ at 4100\,\AA\ (see Table~\ref{tab:ews}); their relative strengths impart a timescale of star formation, with strong [OII] indicating current activity and H$\delta$ representing more prolonged star formation.  The striking contrast in the stacked spectra hints at differences in the star formation histories of these two populations.  The main sequence group has moderate to strong [OII] emission, relatively deep H$\delta$ absorption, and strong Balmer absorption features.  Based on the spectral classifications of \cite{Dressler99}, these are most likely dust-obscured galaxies undergoing bursty star formation, e(a) galaxies.   They also border on the classification of e(c) galaxies, which experience normal, continuous star formation.  As these main sequence galaxies are coeval with $z\sim1$ GCLASS field, and follow the expected SFR versus stellar mass trend, they have probably been recently accreted from the field, and therefore belong to the infall population. 

On the other hand, the sub-main sequence population has weak [OII] emission and a strong 4000-\AA\ break, marked by a sudden drop at [CaII] K and H and lines.  The shallow H$\delta$ absorption is also indicative of older systems: as massive stars die, the H$\delta$ absorption fades.  The lack of strong Balmer absorption precludes the possibility of poststarburst systems.  The stacked spectrum for this population is more consistent with that of a passive galaxy with only a slight hint of star formation; it is classified as right on the border of an older k-type spectrum and e(c) galaxy, as there is some [OII] emission present.  We list equivalent widths for [OII] and H$\delta$, along with D4000 in Table~\ref{tab:ews}.  

\begin{table}
\begin{center}
\caption{Spectroscopic measurements from the weighted, stacked spectra.}
\label{tab:ews}
\begin{tabular}{ccc}
\hline
\multicolumn{1}{}{} &
\multicolumn{1}{c}{Sub-main sequence} &
\multicolumn{1}{c}{Main sequence} \\
\hline
OII EW .......... & 5.4 & 22.0 \\
H$\delta$ EW ........... & 2.4 & 4.7 \\
D4000 ............. & 1.46 & 1.20 \\
\hline
\end{tabular}
\end{center}
\end{table}

We also note the possibility that galaxies on the sub-main sequence could be AGNs, rather than simply star-forming galaxies with lower star formation activity.  This would also be an interesting explanation as it could imply that the AGN is responsible for quenching the star formation through feedback.  For example, \cite{Page12} have claimed to find evidence for quasar-mode feedback in the form of suppressed star formation rates for the most luminous AGNs at $z=1-3$ (but see \cite{Harrison12} for an alternative result).  In this scenario, we could be witnessing the residual star formation in AGNs that are still actively accreting and just beginning to peak in luminosity.

\section{Discussion}
\label{sec:discussion}
We have analyzed a single $z=0.872$ massive cluster from the GCLASS sample from an infrared perspective, investigating how dusty star-forming galaxies behave in the broader context of all spectroscopically-confirmed cluster members.  We discuss our results here, and merge them with a detailed kinematic analysis of each MIPS member.  This allows us to access the role of environment on star-forming galaxies utilizing their previous exposure to high densities, rather than their instantaneous environment.

The ages and stellar masses (Figure~\ref{fig:d4000}) of the MIPS members compared to both cluster and field galaxies suggests that they have been recently accreted from the field.  The line-of-sight velocity distribution for MIPS members (Figure~\ref{fig:v_hist}) exhibits analogous results, as they hug an outer envelope of velocities consistent with the younger cluster members and what would be expected from an infall population.  MIPS cluster members have been observed to share certain properties with field galaxies in previous studies, but are not simply a parallel population in all cases.  For example, \cite{Kocevski11} determined that although MIPS sources have a spatial distribution within the cluster typical of an infalling population, their spectral properties reveal burstier episodes of star formation than their field counterparts at $z\sim0.9$, perhaps provoked by harassment and mergers during assembly.  While the stacked spectrum of our main-sequence galaxies is consistent with e(a) type galaxies---possibly indicative of bursty star formation---we do not see evidence for enhanced activity compared to the field.  Moreover, we uncover an additional branch of 24\,\um\ members that exhibit older, more quiescent spectral features, suggesting the existence of an environmental quenching mechanism that could occur following a possible (though not definite) initial triggering of star formation.

We find the SSFR of MIPS members to be independent of the projected local cluster environment: they maintain the same level of star formation at all densities and radii (Figure~\ref{fig:ssfr_rad}).  Although this is consistent with previous studies \citep[e.g.,][]{Peng10,Lu12, Muzzin12}, it is surprising given an observed increase in the fraction of star-forming galaxies with decreasing density over various star-formation indicators at $z\lesssim1$, for example, [OII] emission lines \citep{Poggianti08,Muzzin12}; H$\alpha$ luminosity \citep{Sobral11}; and 24\,\um\ flux (\citealp[e.g.,][]{Finn10} and the work presented here as shown by the cyan triangles in Figure~\ref{fig:ssfr_rad}).  Many studies have invoked rapid quenching timescales to explain the disparate trends, such that star-forming galaxies are never observed in an intermediate stage---they are either active or completely quenched.  However, when we investigate the SFRs and SSFRs as a function of stellar mass (Figure~\ref{fig:sfr_mstell}), we discover two distinct star-forming branches: one that is in line with the expected main sequence in the $z\sim1$ field and undergoing obscured star formation, and another that displays depressed levels of star formation for a given mass and has a stacked spectrum consistent an older k-type galaxy.  Moreover, these sub-main sequence galaxies are on average more massive; it is the less massive galaxies that are forming the bulk of the stars (though this could be a selection effect).  This is consistent with a downsizing trend in mass assembly such that star formation shifts to less massive galaxies as the Universe ages (see also \citealp{Kodama04, Feulner07, Sobral11}). Perhaps this sub-main sequence is a population of MIPS galaxies that have reached higher density regions or even the cluster core sometime in their past and have been exposed to some sort of environmental quenching; however, this seems inconsistent with the fact that no environmental dependence on SSFR is observed (Figure~\ref{fig:ssfr_rad}).  Either quenching is ubiquitous throughout the cluster with little dependence on the environment or we are not measuring the environment properly.  This raises the question: are we truly sampling MIPS members in the cluster core?  We address this issue in the subsequent sections.

\subsection{Velocity Distributions of Two Star-Forming Populations}
\label{sec:vr_sfr}
\begin{figure*}[]     \centering
\subfigure{\includegraphics[width=9.7cm]{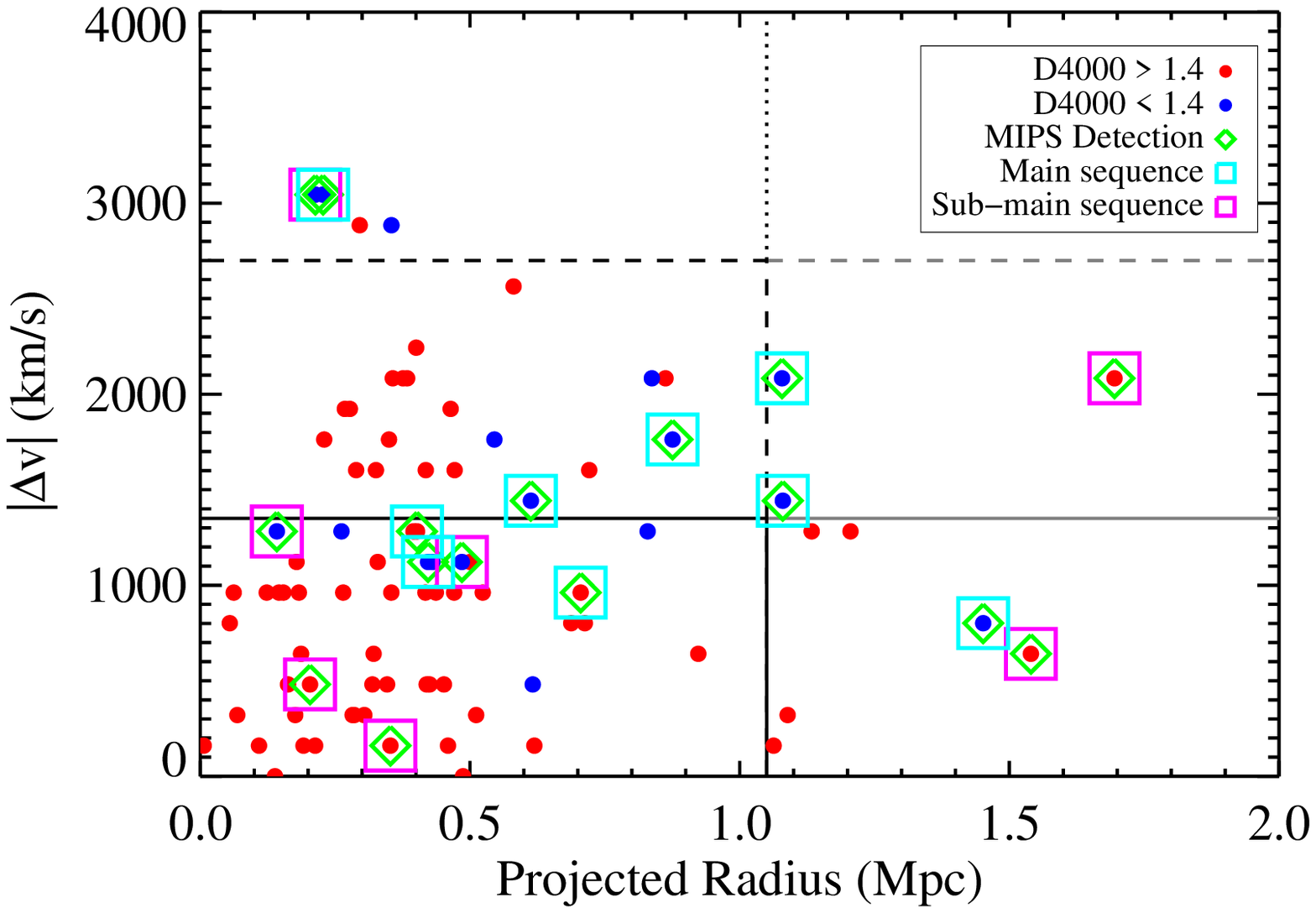} } \hspace{-20mm}
\subfigure{\includegraphics[width=9.7cm]{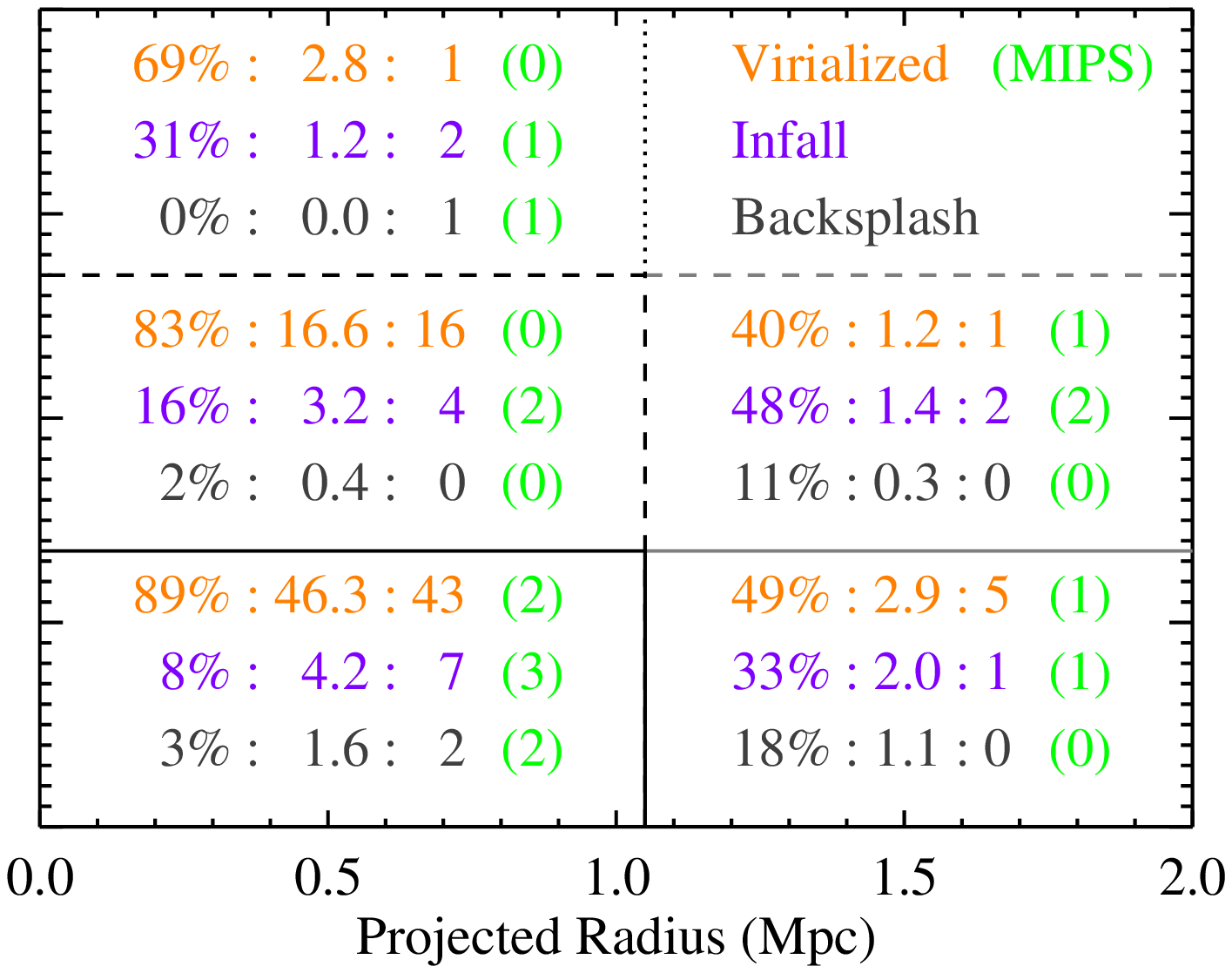} }
   \caption{Left---The absolute line-of-sight velocity as a function of projected radius.  MIPS members belonging to the main (sub-main) sequence of SFR are highlighted with cyan (pink) squares.  The horizontal lines correspond to $\sigma_v$ and $2\sigma_v$, and the vertical line represents $0.5r_{200}$.  The boxed areas correspond to the divisions in \cite{Mahajan11} and are identical in both panels.  Right---The first number on the left in each rectangle corresponds to the fraction of galaxies classified as virialized (orange), infall (purple), and backsplash (gray) in projected phase space bins from dark matter simulations in \cite{Mahajan11}.   We then apply the fractions to the total number of spectroscopically confirmed cluster members in each bin and list it after the colon.  The third number represents the actual number of each type of galaxy within that phase space based on its age and SFR (for MIPS-detected members).  We assume a simple classification in which: virialized galaxies consist of non-star forming older galaxies (red circles) and MIPS-detected older galaxies on the sub-main sequence (red circles with pink squares);  infalling galaxies include non-star forming younger galaxies (blue circles) and MIPS-detected galaxies on the main sequence, both young (blue circles with cyan squares) and old (red circles with cyan squares); and backsplash galaxies comprise younger MIPS-detected galaxies on the sub-main sequence (blue circles with pink squares).  The last column in each bin (green number in parentheses) represents the number of MIPS sources of that galaxy type that are included in the actual number counts based on the classification scheme stated above.} 
   \label{fig:v_sfr}
\end{figure*}

In Figure~\ref{fig:v_sfr} we re-analyze the line-of-sight velocities as a function of projected radius for the MIPS members, accounting for the double-sequence in SFRs.  We highlight the MIPS sources on the main sequence with cyan squares, and those populating the sub-main sequence with pink squares.  There seems to be a rather clear distinction between the two populations, with the majority of sub-main sequence galaxies possessing lower velocities, while the main sequence galaxies avoid velocities below 1000\,km\,s$^{-1}$; with a few exceptions, the main sequence MIPS members adhere to an outer envelope in radial-velocity space.  More importantly, at any given projected radius, we observe a mix of main sequence and sub-main sequence star-forming galaxies.  If we return to the notion that the SSFR has no significant dependence on environment, this result offers a hint of an alternative interpretation: it indicates that low radial bins are contaminated by high-velocity galaxies, which could be infalling galaxies (outside the cluster core) that have fallen far enough into the gravitational potential well to attain large line-of-sight velocities \citep{Haines12}.  Some of these systems may in fact be at physically low radii, however, they could be on their initial (or even a subsequent) pass through the cluster and therefore are still distinct from a virialized core population that has spent a much greater amount of time in dense environments.  As these systems have retained a nominal level of star formation for their stellar mass, they augment the average SSFR in low radial bins, or in what is referred to as the cluster core.  The antithesis contamination occurs at large projected radii where the average SSFR is pulled down from low velocity interlopers---possibly galaxies at large physical radii that have already felt the effects of the cluster environment and therefore are not representative of an infalling population.  In tandem, these two effects yield an overall flat SSFR with projected radii and densities.

We attempt to identify dynamically distinct populations in Figure~\ref{fig:v_sfr} utilizing the results of \cite{Mahajan11}, where they use the dark matter hydrodynamical simulations from \cite{Borgani04} to extract a sample of 93 mock clusters from simulations of 117 haloes, 105 of which have M$_{200}>1.4\times10^{14}\,$\Msol.  From these clusters, they statistically quantify the fraction of virialized, backsplash, and infall galaxies at various projected radii and velocities.  These classifications correspond to decreasing timescales of accretion, respectively: virialized galaxies were accreted when the cluster core was forming and are now passively evolving; backsplash galaxies have had time to pass through the cluster core and experience the first effects of the cluster environment, but have not yet amalgamated with the virialized population and therefore represent an intermediate phase of accretion \citep[e.g.,][]{Balogh00}; infall galaxies were recently accreted from the field and have not reached pericenter in their orbit around the cluster. We plot the delineations in velocity-radius phase space in Figure~\ref{fig:v_sfr}; the corresponding fractions of each classification, along with the applications to our entire spectroscopic sample are summarized in the right-hand panel.

We assign each galaxy in our spectroscopic sample as virialized, infalling, or backsplash based on its age and, for the case of MIPS-detected members, its SFR.  For cluster members that are not detected by MIPS, we simply assume that older members are virialized and younger galaxies have been recently accreted (i.e., infalling).  For MIPS galaxies, we assume any galaxy on the main sequence belongs to the infalling population, while those on the sub-main sequence are virialized if they are older, or are part of the backsplash population if they are younger.  

The consistency between the simulated fractions from \cite{Mahajan11} and our cluster sample, using this simplistic criteria, is quite good.   For example, the solid outlined box in Figure~\ref{fig:v_sfr} represents galaxies within $0.5r_{200}$ and $1.0\sigma_v$, where 89\%, 8\%, and 3\% of galaxies are expected to belong to the virialized, infall, and backsplash populations, respectively; we have 52 cluster members within this area, therefore the predicted fractions correspond to 46.3 virialized, 4.2 infalling, and 1.6 backsplash members in our sample (the second column in the right panel displays the expected number within our sample in each box).  Of the seven MIPS-detected galaxies within this region, three fall on the star-forming main sequence (cyan squares), which suggests they are recently accreted members that have retained their gas and star formation---this number is just below the expected number of infalling galaxies from the simulations (4.2), and increases to seven after we include the members not detected by MIPS. The remaining MIPS members in this region lie on the sub-main sequence (pink squares), and therefore display lower SFRs.  Additionally, there is a clear divide in the age of these four galaxies, with two containing shallow 4000\,\AA\ breaks (filled blue circles).  The depressed SFRs of these two galaxies, along with their younger stellar populations, suggest they are backsplash galaxies which were stripped of their gas reservoir as they passed through the cluster core, but have not had enough time to completely virialize; this is also consistent with the expected backsplash fraction (two galaxies in total).  The last two MIPS members (pink squares surrounding red circles) are both older and have suppressed SFRs, suggestive of passive, virialized galaxies which statistically comprise the bulk of this region.  Including the galaxies without MIPS detections that have older ages, our classification scheme yields 43 virialized galaxies in this region, which is consistent with the number from the simulations (46.3).   We note there is (seemingly) one MIPS outlier at both large projected radius and velocity that has a low SFR and an older stellar population.  However, this region is still expected to contain 40\% virialized galaxies, so it is not unfeasible for less active systems to inhabit this phase space.  

We check the robustness of our classifications with respect to the depth of the MIPS data by assuming the SFR-limit is a factor of two lower (3\,\myr), and all galaxies currently not detected by MIPS have this SFR.  We then simplistically assign them to the above classifications based on their stellar mass and which star-forming sequence they populate.  We find that two infalling galaxies would switch to a backsplash classification and three virialized galaxies would be infalling; therefore, there is no significant change in the numbers assuming a deeper SFR-limit.

\subsection{Utilizing Caustic Diagrams as Accretion History Predictors}
\label{sec:sims}

The mixing of different dynamical histories at all projected radii casts a doubt on the interpretation that environment has no effect on the SSFR of star-forming galaxies.  However, our attempt to classify each galaxy as infalling or belonging to the cluster core is circumstantial.  

A complementary method to kinematically differentiate between galaxies arises with the use of caustic profiles (i.e., constant velocity-radial lines) in physical radial phase space, where infalling and virialized cluster galaxies become distinct \citep[e.g.,][]{Mamon04, Gill05, Mahajan11, Haines12}.  Moreover, this is inherently linked to the epoch of accretion, which illustrates the importance of a dynamically-defined environment: the time-averaged density a galaxy has been subjected to is more significant than its immediate surroundings.  

Recently, \cite{Haines12}, stacked 30 clusters in the Millennium Simulation at $z=0.21$ and investigated the relation between projected measurables and the accretion history of the cluster (see fig.~3 in \citealp{Haines12}).  Galaxies that were accreted when the cluster was first forming have a large dynamic range of velocities at the smallest projected radii, but only occupy a narrow silver of velocities ($\Delta v/v_\sigma\simeq0$) over larger projected radii.  Conversely, recently accreted galaxies permeate all projected velocities and radii, but are primarily concentrated along and outside a caustic line.  These trumpet-shaped caustic profiles roughly correspond to lines of constant $r\times v$ in projected phase space.  Recently accreted galaxies are therefore the dominant contaminant compared to galaxies that were accreted earlier, specifically in low radial bins.  Moreover, star forming studies are biased towards picking out these contaminants, assuming recently accreted galaxies have managed to retain their activity and are classified as star-forming.  This could explain why SSFR trends utilizing the entire galaxy population (quiescent and star-forming) or stacked SSFRs preserve a correlation with environment \citep[e.g.,][]{Patel11}, while those solely for star-forming galaxies flatten out.

\begin{figure}[h!]    
 \centering
\includegraphics[width=9cm]{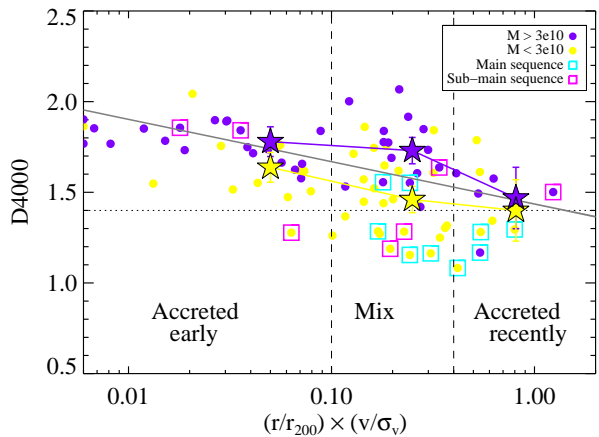}
   \caption{The strength of the 4000-\AA\ break versus $(r/r_{200})\times(\Delta v/\sigma_v)$ for all cluster members, divided in two mass bins.  The MIPS members on the main and sub-main sequence are highlighted with cyan and pink squares, respectively. The solid gray line shows the best fit line for all galaxies.  The median D4000 values for both mass bins in each region (denoted by the vertical dashed lines) are plotted as filled stars and indicate a declining accretionary sequence with $(r/r_{200})\times(\Delta v/\sigma_v)$. The error bars are taken from the standard deviation on the bootstrapped median in each bin.  We have labeled the regions based on accretion times.  We note the three galaxies with the lowest caustic values are artificially plotted at $(r/r_{200})\times(\Delta v/\sigma_v)=0.006$ to reduce the size of the plot, but in fact have even lower values.}
   \label{fig:d4000_vr}
\end{figure}

Motivated by the possibility of identifying distinct accretion histories within our own sample via caustic profile diagnostics, in Figure~\ref{fig:d4000_vr} we plot the strength of the 4000-\AA\ break as a function of $(r/r_{200})\times(\Delta v/\sigma_v)$, since constant values of this parameter correspond to caustic profiles which trace out accretion epochs.  There is a clear movement towards higher values of $(r/r_{200})\times(\Delta v/\sigma_v)$ for shallower breaks, meaning lower values on the x-axis correspond to older galaxies that were accreted at earlier times.  We calculate a linear Pearson correlation coefficient of $-0.41$ between the parameters, with a high significance of $>99.99\%$. To ensure this trend is not driven purely by the correlation between D4000 and stellar mass, we have separated the cluster members into two mass bins, above and below $3\times10^{10}$\,\Msol, which is roughly the median mass of the sample.  The relation persists for both mass bins, meaning that for cluster galaxies of the same stellar mass, there is a progression towards larger $(r/r_{200})\times(\Delta v/\sigma_v)$ values with decreasing time since the last burst of star formation.  Assuming higher values of $(r/r_{200})\times(\Delta v/\sigma_v)$ trace galaxies that were more recently accreted (which is analogous to fig.~3 in \citealp{Haines12}), the trend in Figure~\ref{fig:d4000_vr} indicates the cluster environment has an effect on its constituent galaxies.   

In order to isolate contamination in low projected radial bins from infalling/backsplash interlopers, we divide the space into three distinct regions based on the location of main and sub-main sequence galaxies.  Combining the results of \cite{Haines12} and \cite{Mahajan11}, these regions should primarily correspond to: virialized galaxies that were accreted when the cluster core was forming; a mix of all types but where backsplash galaxies are most likely to exist; and infalling galaxies that were recently accreted---at low, intermediate, and high values of $(r/r_{200})\times(\Delta v/\sigma_v)$, respectively.  We calculate the median D4000 in each region for both mass bins (filled stars), which confirms the declining age (i.e., epoch of accretion) with $(r/r_{200})\times(\Delta v/\sigma_v)$.

\begin{figure}[h!]     \centering
\includegraphics[width=9cm]{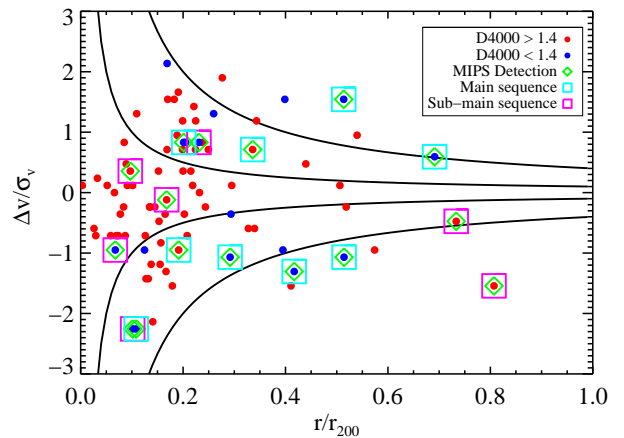}
 \caption{A caustic diagram for all spectroscopically confirmed cluster members.  The symbols are the same as in Figure~\ref{fig:v_sfr}.  We plot caustic profiles---lines of constant $(r/r_{200})\times(\Delta v/\sigma_v)$---at 0.1 and 0.4 to isolate virialized (within the inner caustic), backsplash (between caustics), and infall galaxies (along and outside the outer caustic).}
   \label{fig:vr}
\end{figure}

We demonstrate the utility of these caustic regions in reference to our own sample in Figure~\ref{fig:vr}.  We plot lines of constant $r\times v$ at 0.1 and 0.4 ($r_{200}\times\sigma_v$), based on the regions in Figure~\ref{fig:d4000_vr}.  The innermost caustic line seems to preferentially select the older (virialized) population of cluster members that were most likely accreted at early times, including three out of the seven MIPS sources belonging to the sub-main sequence branch.  Although rare, it is not inconceivable that recent star formation has occurred within the virialized population; \citep{Mahajan11} estimate that 11\% of the virialized population are galaxies with ongoing or recent efficient star formation that could be attributable to rapid flybys.  However, we note that their level of star formation is certainly suppressed compared to the main sequence.

Between the two caustic profiles, we expect a distribution of galaxies that were accreted early, recently, and somewhere in-between.  Indeed, in this intermediate region of $0.1<r/r_{200}\times v/\sigma_v<0.4$, there exists a mix of old, main sequence galaxies and young, sub-main sequence galaxies, which could result from a backsplash population. The significance of the backsplash population has been demonstrated before in various simulations.  \cite{Balogh00} found that over half of the galaxies within 1--2$r_{200}$ have been inside the virial radius at an earlier time and rebounded outward.  Moreover, 90\% of these backsplash galaxies have plunged deep into the potential well, within the inner 50\% of the virial radius \citep{Gill05}.  

The region exterior to the 0.4 ($r_{200}\times\sigma_v$) caustic profile should preferentially pick out galaxies that have been recently accreted \citep{Haines12}.  Based on the fractions from \cite{Mahajan11}, it should additionally favor infalling over backsplash galaxies.  Moreover, \cite{Gill05} observed that basksplash galaxies are kinematically distinct from infalling galaxies, as the latter have much higher relative velocities.  Although this velocity bimodality can get slightly washed out in projected space, there still exists an average trend toward lower velocities for backsplash galaxies.  Notably, this region contains the majority of young main sequence galaxies, which are most consistent with an infall population given their age and SFRs.  As stated in \S\ref{sec:vr_sfr}, the lone sub-main sequence galaxy in this region could statistically be a virialized galaxy outside the core; however, we also note that we expect one chance alignment between MIPS sources and spectroscopic members (\S\ref{sec:false}), which could be manifested here.  Alternatively, this galaxy and another old, sub-main sequence galaxy at high projected radius could both belong to groups that have been accreted into the cluster.  In this scenario, the lower SFRs of these galaxies could be due to pre-processing \citep[e.g.,][]{Zabludoff98} in the group environment prior to accretion.  \cite{McGee09} found that the accretion of galaxy groups onto a massive $z\sim1$ cluster accounts for a significant proportion ($\gtrsim$30\%) of the galaxy population, and moreover, galaxies that derive from group environments are more massive.  Indeed, the two outliers are both more massive than the majority of main sequence galaxies and exhibit suppressed SFRs.  However, the sparse spectroscopy at these high radii precludes us from reliably identifying infalling groups.

\subsection{A Kinematic Approach to the Environmental Dependence on the Specific Star Formation Rate}
\label{sec:ssfr_caustic}
\begin{figure}[h!]     \centering
\includegraphics[width=9cm]{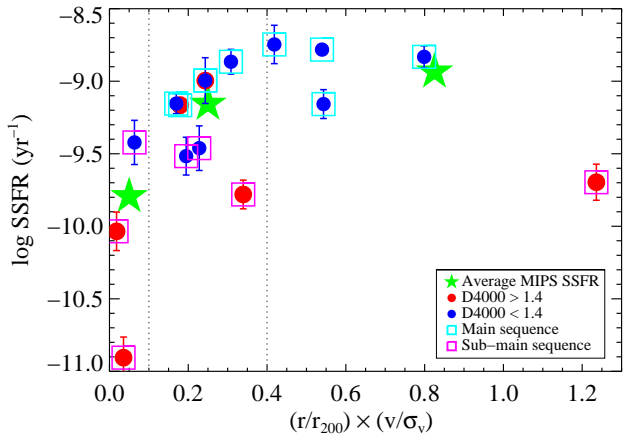}
   \caption{The SSFR for star-forming galaxies as a function of caustic environment, $(r/r_{200})\times(\Delta v/\sigma_v)$.  The sub-main sequence galaxies (pink squares) primarily inhabit low values of $(r/r_{200})\times(\Delta v/\sigma_v)$ compare to the main-sequence galaxies.  In contrast to Figure~\ref{fig:ssfr_rad}, there is now a clear depression (0.9 dex) of the average SSFR (green stars) towards low caustic values, where we expect to isolate galaxies accreted at earlier times.  The gray vertical lines indicate the limits of the bins.}
   \label{fig:ssfr_vr}
\end{figure}

In hopes of disentangling radial contamination from environment, we consider the SSFR as a function of $r\times v$, normalized by the cluster velocity dispersion and $r_{200}$, in Figure~\ref{fig:ssfr_vr}.  As shown in Figures~\ref{fig:d4000_vr}, \ref{fig:vr}, and fig.~3 from \cite{Haines12}, low values of $r\times v$, $<0.1 (r_{200}\times\sigma_v)$, should represent a more dynamically-isolated cluster core, as it is dominated by galaxies accreted at early times, which have spent more time in dense cluster regions.  Similarly, galaxies recently accreted possess higher caustic values of $>0.4 (r_{200}\times\sigma_v$), as they are more likely to have high projected radii and velocities.  In Figure~\ref{fig:ssfr_vr} there is a clear segregation between the two star forming populations when using a caustic approach to define environment.  The sub-main sequence galaxies (pink squares) are preferentially at low values of $r\times v$ and well contained within 0.4 ($r_{200}\times\sigma_v$) with the exception of one galaxy; the main-sequence branch (cyan squares), however, does not penetrate below a value of 0.1 ($r_{200}\times\sigma_v$), but rather is dispersed throughout intermediate and high values.  The isolation of the galaxies on the sub-main sequence yields a prominent depression in SSFR towards the ``caustic core" of the cluster, namely, with galaxies that exhibit both low projected radii and low line-of-sight velocities. Splitting the galaxies into the aforementioned $r\times v$ bins (shown by vertical lines in Figure~\ref{fig:ssfr_vr}) and calculating the average SSFR of the MIPS cluster members in each bin (green stars) reveals a 0.9 dex decline between the highest and lowest caustic bins.  With this definition for environment, the cluster core is more of a signpost for the dynamic history of its constituents, as the time-averaged density of a galaxy should increase with its time since accretion.  If this dynamic core more accurately represents the virialized cluster core than clustercentric radius or density probes, the interpretation that the SSFR of star-forming galaxies lacks a dependence with environment warrants caution.  Moreover, a rapid quenching timescale would no longer be a necessary requirement; if younger star-forming galaxies on the sub-main sequence are backsplash galaxies, we can use the dynamical timescale to place a lower limit of $\sim$1.3\,Gyr on the duration of quenching.

\section{Conclusions}
\label{sec:conclusions}
We have presented an infrared study of SpARCS J161314+564930, a $z=0.872$ massive galaxy cluster comprising 85 mass-limited, spectroscopically confirmed members, 16 of which have been detected at 24\,\um.  After combining the stellar ages, SFRs/SSFRs, and kinematics of dusty star forming galaxies, a clean narrative of the history of these cluster members begins to unfold.  We summarize the main properties and trends of the MIPS galaxies as follows:
\begin{enumerate}

\item{MIPS cluster members form a young envelope of the cluster population; they span the same range of stellar masses and D4000 values as coeval field galaxies, suggesting they primarily belong to a recently accreted population.  }

\item{There is a bimodal or flat distribution of velocities for the MIPS members, peaking at $\pm1000$\,km\,s$^{-1}$, which is roughly the cluster velocity dispersion.  This sharply deviates from the older cluster members, which form the expected Gaussian distribution of velocities.  There is only a 6\% probability that these two distributions derive from the same parent population; therefore, the MIPS sources seem to represent a recently accreted, un-virialized population.  }

\item{When local environment is defined as projected clustercentric radius or density, it appears to have no effect on the average SSFR of star-forming galaxies (but see conclusions 5--8 for an alternative explanation).  However, when accounting for the total mass from all cluster members, there is a sharp decline in the total SSFR towards low radii, indicating that the SFR-density relation is already established in the densest regions at $z\sim0.9$.}

\item{There exist two branches of SFR/SSFR as a function of stellar mass---one that is consistent with the star-forming main sequence of increasing SFR with mass, while the other displays a flat trend with suppressed SFRs compared to the $z\sim0.9$ field. This double-sequenced distribution appears to be unique to the cluster. Stacking the spectra of galaxies along each branch separately supports this by revealing distinct star formation histories: the former are likely obscured galaxies experiencing either continuous or bursty  star formation, while the latter lacks Balmer absorption, has a strong 4000-\AA\ break, and weak [OII] emission---features that are more typical of passive or quiescent galaxies.}

\item{Utilizing the results from \cite{Haines12}, we show that recently accreted galaxies contaminate all projected radii, while the earliest accreted galaxies are primarily at low radii, with only a little spillover at larger radii.  Assuming star forming galaxies are preferentially an infalling or recently accreted population, the star forming population is therefore inherently more likely to be a contaminant in projected environment, which could contribute to the flat trend between the SSFR of star-forming galaxies with environment.  Therefore, low radial bins are contaminated by recently accreted MIPS sources that have not yet (necessarily) passed through the densest regions of the cluster or core and could augment the SSFR.}

\item{A caustic digram of line-of-sight velocity versus projected radius reveals that there is a mix of galaxies from each of the two star-forming sequences at any given radius. We are able to successfully isolate galaxies that were accreted at early times from recently accreted objects via caustic values of $(r/r_{200})\times(\Delta v/\sigma_v)<0.1$ and $>0.4$, respectively. Intermediate values contain a mix of accretion histories, and should also contain the statistical majority of backsplash galaxies.}

\item{Applying caustic profiles of constant  $(r/r_{200})\times(\Delta v/\sigma_v)$ to our sample allows us to kinematically classify MIPS galaxies as virialized, backsplash, or infalling based on their age and SFR.  Galaxies with lower star formation rates (i.e., on the sub-main sequence) appear to belong to either the virialized or backsplash populations.  Further differentiation between these populations arises from stellar age, as measured through the depth of the 4000-\AA\ break.  Although not definitively, older galaxies with D4000 $>$ 1.4, are likely to be virialized within the 0.1 ($r_{200}\times\sigma_v$) caustic profile.  Backsplash galaxies appear to lie in intermediate regions of $0.1<r/r_{200}\times v/\sigma_v<0.4$.  Infalling galaxies primarily occupy the region outside the 0.4 ($r_{200}\times\sigma_v$) caustic profile and have SFRs consistent with the main sequence.}

\item{Using a caustic definition for environment reveals almost an order of magnitude (0.9 dex) depression of SSFR at low $(r/r_{200})\times(\Delta v/\sigma_v)$ values, in contrast to the flat correlation observed with projected radius or density.  This suggests that environment may suppress star formation as galaxies fall deeper into the cluster potential, with a minimum quenching timescale given by the infall time.  }

\end{enumerate}

We stress that this is a case study of a single galaxy cluster, and a larger sample is needed to statistically verify our conclusions.  We plan to expand our study to the entire GCLASS sample, which consists of ten clusters over $0.87<z<1.35$ and over 400 confirmed members.  With a significant sample of dusty star-forming galaxies we can better assess the diagnostic power of caustic profiles in relation to environment, stellar ages, star formation, and quenching mechanisms.

\acknowledgments
We thank the anonymous referee for helpful suggestions which improved the manuscript.  T.M.A.W. gratefully acknowledges the support of the NSERC Discovery Grant and the FQRNT Nouveaux Chercheurs program.  G.W. acknowledges support from NSF grant AST-0909198.  H.K.C.Y. is supported by the NSERC Discovery Grant and a Tier 1 Canada Research Chair. R.F.J.vdB. acknowledges support from the Netherlands Organization for Scientific Research grant number 639.042.814.

\textit{Facilities:} \facility{Spitzer Space Telescope (MIPS; IRAC)}, \facility{Gemini (GMOS)}

\bibliography{references}
\bibliographystyle{apj}

\end{document}